\documentclass{aa}  

\usepackage{float}
\usepackage[varg]{txfonts}
\usepackage{amssymb}
\usepackage{graphicx}
\newcommand\degree{\ensuremath{^\circ}}
\usepackage{graphicx}	
\usepackage{mathtools}	
\usepackage{amssymb}	
\usepackage{multicol}        
\usepackage{bm}		
\usepackage{pdflscape}	
\usepackage{xcolor}
\usepackage{placeins}
\usepackage{txfonts}
\begin{document} 

   \title{Classical, large scale 3D MHD simulations \\ of interacting pulsar 
   wind nebulae}


   \author{D. M.-A.~Meyer\inst{1}
          and 
          D. F. Torres\inst{1,2,3}\\           
          }

    \institute{Institute of Space Sciences (ICE, CSIC), Campus UAB, Carrer de Can Magrans s/n, 
    08193 Barcelona, Spain \\
    \email{meyer@ice.csic.es ; dmameyer.astro@gmail.com}
    \and
    Institut d’Estudis Espacials de Catalunya (IEEC), 08034 Barcelona, Spain  \\
    \email{dtorres@ice.csic.es}    
    \and
    Institució Catalana de Recerca i Estudis Avançats (ICREA), 08010 Barcelona, Spain    
          }
   \date{}


  \abstract 
   {  
Magnetized rotating neutron stars, or pulsars, are a possible end product of massive star 
evolution. Their relativistic wind successively interacts with the supernova ejecta of their 
defunct progenitor, then with the circumstellar medium of the progenitor, and eventually with 
the interstellar medium. The distribution of those materials governs the morphology, mixing of 
chemical elements, and emission properties of the shocks present in plerionic supernova remnants. 
   }
   {     
If a massive star is static with respect to its ambient medium, then its resulting circumstellar 
medium is elongated along the direction of the local magnetic field, and its supernova remnant transiently appears as a rectangle. The pulsar wind nebula forming in it is, in its turn, 
elongated, as long as the pulsar's axis of rotation matches the direction of the local magnetization. In this work, we explore how the angle between the direction of the local magnetic field of 
the interstellar medium and the pulsar axis of rotation influences the shaping of its pulsar 
wind nebula. 
   }
   { 
 Three-dimensional magneto-hydrodynamic simulations are carried out with the PLUTO code to 
 model the pulsar wind nebula formed by a static pulsar inside of a supernova remnant left 
 behind by a massive Wolf-Rayet-evolving progenitor at rest in an organized, magnetized ambient 
 medium. We use those models to perform radiative transfer calculations to derive non-thermal 
 radio emission maps of the pulsar wind nebulae. 
   }
   { 
When the \textcolor{black}{polar elongation} of the pulsar develop, 
they bend in opposite directions under the effects 
of the cavity carved by the stellar wind and already filled by supernova ejecta. This induces 
a complex distribution of magnetized supernova ejecta and pulsar wind, resulting in various 
observable structures, appearing as rectangles, circles, or irregular oblong shapes, in the 
radio waveband. 
   }
   {  
The angle between the direction of the pulsar rotation axis and that of the local ambient 
magnetization is a governing parameter for the shaping and non-thermal radio properties of 
the pulsar wind nebulae of static massive stars; however, the mixing of material, once the 
pulsar wind nebula is old ($50$$-$$80\, \rm kyr$), is not strongly affected by that factor. 
   }


   \maketitle
%


\section{Introduction}
\label{intro}

Pulsars are highly magnetized, rotating neutron stars formed from the explosive 
deaths of massive stars, provided the collapse does not directly result in a 
black hole~\citep{Gold1968,Pacini1968,Lyne2012}. 
Their magnetized surfaces launch powerful relativistic winds, which interact 
with the surrounding environment to produce pulsar wind nebulae, as reviewed 
in~\citet{Gaensler_Slane_2006ARA&A..44...17G,Kargaltsev_etal_2017JPlPh..83e6301K, 
Olmi_Bucciantini_2023PASA...40....7O}.
These nebulae emit across the entire electromagnetic spectrum, with synchrotron 
radiation and inverse Compton scattering dominating their non-thermal emission 
\citep{kennel_apj_283_1984,Gaensler_Slane_2006ARA&A..44...17G}. 
Additional features include distinct emission 
lines~\citep{Reynolds_etal_2017SSRv..207..175R}, as well 
as observational signatures in the radio band (e.g., G54.1+0.3~\citep{Driessen_etal_2018ApJ...860..133D}), 
X-rays (e.g., G11.2$-$0.3~\citep{Borkowski_etal_2016ApJ...819..160B}), and 
gamma-rays extending to PeV energies, notably in the Crab Nebula~\citep{2006A&A...457..899A,
Abdall_etal_2021ApJ...917....6A, Lhaaso_Collaboration_2021Sci...373..425L, acero_aph_2023}.
The morphology and dynamical evolution of pulsar wind nebulae 
are significantly shaped by their 
surrounding medium, whether confined within progenitor supernova remnants or 
propagating through the interstellar medium (ISM) as bow shocks, following the 
natal kicks imparted to pulsars~\citep{temim_apj_808_2015}. 
The interaction between pulsar winds and the dense, anisotropic ejecta of 
core-collapse supernovae can generate complex structures and irregular emission
features~\citep{Reynolds_etal_2017SSRv..207..175R,Olmi_Bucciantini_2023PASA...40....7O}. 
Comprehensive catalogues of observed pulsar wind nebulae have been assembled 
in multiple studies~\citep{Hester_2008ARA&A..46..127H,behler_RPPh_2014,bock_aj_1998, 
Kargaltsev_apss_308_2007,Kargaltsev_apjs_201_2012,2017JPlPh..83e6301K, 
frail_apj_480_1997,hess_aa_612_2018,popov_2019,2019A&A...627A.100H, 
acero_aph_2023,turner_mnras_531_2024}.

Pulsar wind nebulae are extreme environments that can only be effectively probed through 
numerical simulations. The body of literature on simulations of pulsar wind nebulae 
generally divides into two main groups of studies.
On the one hand, there are models focusing on the immediate surroundings of the pulsar 
and tracking the evolution of the nebula up to timescales of approximately $1\, \rm kyr$ 
after the onset of the pulsar wind \citep{kennel_apj_283_1984,coroniti_apj_349_1990,
begelman_apj_397_1992,begelman_apj_493_1998,komissarov_mnras_344L_2003,swaluw_aa_397_2003,
komissarov_mnras_349_2004,swaluw_aa_420_2004,komissarov_mnras_367_2006,
Del_Zanna_etal_2006A&A...453..621D,Camus_etal_2009MNRAS.400.1241C,
komissarov_mnras_414_2011,Olmi_etal_2014MNRAS.438.1518O,Porth_etal_2014MNRAS.438..278P,
Olmi_etla_2016JPlPh..82f6301O}. 
On the other hand, there are simulations that encompass the entire circumstellar medium 
and the supernova ejecta, extending to about $30\, \rm kyr$ after the explosion 
\citep{temim_apj_808_2015,temim_apj_851_2017,kolb_apj_844_2017,temim_apj_932_2022}, 
or tracking the nebula as it moves through the ISM after being expelled from 
\textcolor{black}{
its parent 
supernova remnant \citep{Bucciantini_aa_375_2001,bucciantini_mnras_478_2018,
barbov_mnras_2018,barkov_mnras_484_2019,barkov_mnras_485_2019,
Toropina_etal_2019MNRAS.484.1475T,olmi_MNRAS_484_2019,olmi_MNRAS_488_2019,
olmi_MNRAS_490_2019}. 
}
Among the various factors influencing the morphology of pulsar wind nebulae, the 
interaction of the progenitor's stellar winds plays a dominant role in shaping 
their structure and emission properties~\citep{meyer_mnras_515_2022,
meyer_mnras_537_2025,2025A&A...696L...9M}.

The circumstellar medium of massive stars is commonly described by the stellar wind 
bubble model~\citep{weaver_apj_218_1977}. This model envisions the stellar surroundings 
as a series of shocks and discontinuities, bounded by an inner termination shock and 
an outer forward shock. Within this structure, concentric and often unstable layers 
of hot, diluted gas and cold, dense stellar wind material form.
Wind bubbles are influenced by several factors, including the bulk motion of the 
central massive star, which can transform the circumstellar medium into a stellar 
wind bow shock \citep{gull_apj_230_1979,wilkin_459_apj_1996}, stellar rotation, 
and the granularity of the multi-phase ambient medium~\citep{mckee_apj_195_1975}. 
The circumstellar medium is continuously reshaped by the stellar feedback from its 
central star. Regular variations in the stellar wind, which mark the transition between 
different evolutionary phases following the long initial main-sequence stage, further 
modify the internal regions of the circumstellar medium in terms of momentum and 
chemical composition. This evolution gives rise to structures associated with 
red supergiants~\citep{noriegacrespo_aj_114_1997,decin_paper_Aori} or luminous 
blue variables~\citep{gonzales_aa_561_2014}.
At the end of the star's life, the supernova blast wave interacts with the 
pre-shaped circumstellar medium, generating a supernova remnant whose properties 
reflect the evolutionary history of its progenitor~\citep{weiler_aa_70_1978,
caswell_mnras_187_1979,weiler_aa_90_1980,Gaensler_Slane_2006ARA&A..44...17G,
Kargaltsev_etal_2017JPlPh..83e6301K}.

The influence of the magnetic field in the ISM on the morphology 
of circumstellar nebulae is well-studied. It serves as a stabilizing mechanism for the 
instabilities of stellar wind bow shocks from massive runaway 
stars~\citep{meyer_mnras_464_2017,baalmann_aa_650_2021,meyer_mnras_506_2021}, elongates 
the shapes of stellar wind bubble and superbubbles~\citep{vanmarle_584_aa_2015} along 
the direction of the local ambient magnetic field. 
After the explosive death of massive progenitor stars, their magnetized circumstellar 
medium presents an obstacle to the expanding supernova 
blastwave~\citep{langer_araa_50_2012,orlando_aa_622_2019,orland_aa_636_2020,
orlando_aa_645_2021,orlando_aa_666_2022,2025arXiv250314455O}. 
This interaction can result in an asymmetric supernova remnant~\citep{meyer_mnras_502_2021}, 
modify the mixing of the ejecta with surrounding ISM material~\citep{meyer_mnras_521_2023}, 
and influence phenomena such as non-thermal emission~\citep{meyer_2024_loops}.

About a third of massive stars, i.e. of core-collase supernova progenitors, move supersonically 
through the
\textcolor{black}{
ISM~\citep{wit_aa_437_2005} and generate a dense stellar wind 
bow shock~\citep{vanburen_aj_110_1995} 
}
which strongly affects the propagation 
of the supernova blastwave~\citep{meyer_mnras_450_2015}. 
We have investigated the scenario of a young pulsar 
in such supernova remnant in~\citet{2025A&A...696L...9M}. 
In the case of a static supernova progenitor, the reflection of the supernova blastwave 
off the walls of the oblong cavity generates a tubular structure, which appears as a 
rectangle in projected emission~\citep{2022MNRAS.515..594M} and notable examples of such 
objects include the supernova remnant Puppis A. 
As demonstrated in previous studies, the morphology of pulsar 
wind nebulae originating from massive progenitors is also influenced by the local 
distribution of ejecta, stellar wind, and ISM. Consequently, the pulsar wind nebulae 
of static massive progenitors are stretched along the direction of the ISM's magnetic 
field~\citep{meyer_527_mnras_2024}.

The two-dimensional nature of the models in~\citet{meyer_527_mnras_2024} 
obliged us to consider colinear directions 
for both the ISM magnetic field and the axis of rotation of the pulsar. 
In this work, we extend these numerical efforts by presenting 3D MHD simulations of the pulsar 
wind nebula of a static massive star. The effects of the magnetization of the ISM on the elongation 
of its progenitor's circumstellar medium at the pre-supernova time are included, and we present 
non-thermal radio emission maps of our results. In Section~\ref{method}, we present the methods 
used in this study; in Section~\ref{results}, we detail the obtained results, which are further 
discussed in Section~\ref{discussion}. We draw our conclusions in Section~\ref{conclusion}.


\section{Method}
\label{method}

\subsection{Simulation strategy}
\label{method_strategy}

In this study, we perform a series of global simulations for the pulsar wind nebulae 
of a static massive star. The models are carried out within the framework of classical 
magnetohydrodynamics using the {\sc PLUTO} 
code~\citep{mignone_apj_170_2007,migmone_apjs_198_2012}. 

Since the circumstellar medium of massive stars governs the morphology and various 
properties of supernova remnants, and, through density wave reflections, it also influences 
the pulsar wind nebulae forming in the relics of core-collapse supernovae, our model covers 
the entire evolution of the surroundings of the massive progenitor, from the zero-age 
main-sequence phase up to $80\, \rm kyr$ after the supernova explosion. 
First, the circumstellar medium, produced by the interaction of the massive star’s stellar 
wind with the ISM, is modeled throughout the star's entire life. It is simulated in $2.5$D, 
using a cylindrical coordinate system (2D for scalar quantities plus a toroidal component 
for the vectors), in the frame of the static star, to be used as initial conditions for 
the supernova remnant phase calculation~\citep{2025A&A...696L...9M}. 
Secondly, the interaction of the supernova blast wave with the freely-expanding stellar 
wind from the progenitor's last evolutionary phase is modeled in a 1D spherical coordinate 
system until its forward shock reaches a distance of $2\, \rm pc$, smaller than the termination 
shock of the stellar wind bubble. 
Thirdly, a 3D Cartesian simulation is performed, using the $2.5$D circumstellar medium 
that is rotated around its symmetry axis to reconstruct a 3D structure. The 1D solution 
is then mapped onto this 3D structure, assuming an isotropic supernova explosion, 
providing the initial conditions for the global plerionic supernova remnant models. 
This study focuses on pulsar wind nebulae which are interacting with their pre-supernova 
circumstellar medium, and, on the associated aftermaths of it. In the warm phase of the 
ISM, this corresponds to older plerion ($\ge 40\, \rm kyr$), see ~\citet{meyer_527_mnras_2024}. 
The workflow is summarised in Fig. \ref{fig:concept_plot}.

\subsection{Interstellar medium}
\label{method_ism}

The single massive star considered in this study is assumed to reside in a uniform medium 
that mimics the plane of the Milky Way, specifically in the warm phase of the ISM. Its temperature 
is set to $T_{\rm ISM} = 8000\, \rm K$, with a number density of $n_{\rm ISM} = 0.79\, \rm cm^{-3}$ 
and a magnetic field strength of $B_{\rm ISM} = 7\, \mu\, \rm G$
\citep{wolfire_apj_587_2003,meyer_2024_loops}.
The ISM is assumed to be in equilibrium with optically-thin radiative processes for cooling 
and heating. Coolants from elements such as H, He, and metals (at solar He abundance) are 
included, following the prescriptions of~\citet{wiersma_mnras_393_2009,asplund_araa_47_2009}. 
Additionally, our model incorporates contributions from H recombination 
lines~\citep{hummer_mnras_268_1994}, as well as emission from forbidden lines of O and 
C~\citep{henney_mnras_398_2009}. Cooling by O is particularly important for shock physics, 
and it is assumed to be present in the ISM at an abundance of 
$\rm O/\rm H = 4.89 \times 10^{-4}$ in number density, as presented 
by~\citet{asplund_araa_47_2009}. Starlight heating, through the recombination of 
hydrogenic ions ionized by stellar irradiation, is included as described 
in~\citet{osterbrock_1989}.
The background magnetic field initially filling the ISM is uniform and linearly 
organized, which is consistent with the large scale coherent magnetic fields 
observed in the Galactic spiral arms of the Milky Way~\citep{rand_343_apj_1989}. 
\textcolor{black}{
Since the circumstellar medium of the massive star is modeled in two dimensions, 
it forces to adopt the configuration of a runaway progenitor moving 
parallel to the magnetic field lines of the local ambient 
medium, see~\citet{meyer_mnras_464_2017}. 
}

\subsection{Stellar wind interaction with the ambient medium}
\label{method_sw}

The interaction of the stellar wind with the ISM is simulated by placing a $35\, \rm M_{\odot}$ 
static star for which the time-dependent stellar surface properties, namely the mass-loss rate, 
the wind velocity, the effective temperature are taken from the {\sc Geneva} library of 
evolutionary track \citep{ekstroem_aa_537_2012}. The value of the star terminal wind speed 
is calculated from the stellar escape velocity using the 
\textcolor{black}{recipe} of \citet{eldridge_mnras_367_2006}. 
The stellar rotation is selected to be $\Omega_{\star}(t=0) / \Omega_{\rm K} = 0.1$ with 
$\Omega_{\star}(t=0)$ and $\Omega_{\rm K}$ the zero-age main-sequence rotational velocity 
at the equator and the Keplerian rotational speed, respectively. This permits to 
include both the latitude-dependent toroidal component of the stellar rotation in the 
spirit of ~\citet{parker_paj_128_1958,weber_apj_148_1967,chevalier_apj_421_1994,
rozyczka_apj_469_1996,pogolerov_aa_321_1997,pogolerov_aa_354_2000}. 
The stellar magnetic field structure is assumed to be a Parker spiral with a magnetic 
field strength at the surface scaled as the Sun's one ~\citep{scherer_mnras_493_2020,
herbst_apj_897_2020,baalmann_aa_634_2020,baalmann_aa_650_2021,meyer_mnras_506_2021} 
and which value is $B_{\star}=500\, \rm G$~\citep{fossati_aa_574_2015,castro_aa_581_2015}, 
$B_{\star}=0.2\, \rm G$~\citep{vlemmings_aa_394_2002,vlemmings_aa_434_2005} 
and $B_{\star}=100\, \rm G$~\citep{meyer_mnras_507_2021} for the main-sequence, 
red supergiant and Wolf-Rayet phases, respectively. 
The simulation of the CSM surrounding the massive star is performed using a 2.5D 
cylindrical coordinate system. 
It employs a uniform grid covering the range $[0,150] \times [-50,50]\, \rm pc$  
consisting of $7500 \times 5000$ or a uniform spatial resolution of 
$0.02\, \rm pc\, \rm zone^{-1}$.

\subsection{Supernova blastwave interaction with the stellar wind}
\label{method_snr}

The supernova blast wave resulting from the explosion is simulated using a 1D spherically 
symmetric coordinate system, extending to a maximum radius of $r_{\rm max} = 2\, \mathrm{pc}$. 
During this initial phase, the shock front propagates through the freely expanding stellar 
wind, which is assumed to be isotropic. The numerical grid consists of $50000$ uniformly 
spaced zones, providing sufficient resolution to accurately capture the shock dynamics.
The core-collapse supernova explosion model adopted in this study follows the prescriptions of~\citet{whalen_apj_682_2008}, \citet{vanveelen_aa_50_2009}, and~\citet{truelove_apjs_120_1999}.
The total ejected mass is set to $M_{\rm ej} = 10.12\, \mathrm{M_{\odot}}$, obtained by 
subtracting the neutron star mass of $1.4\, \mathrm{M_{\odot}}$ from the progenitor’s 
final stellar mass at the time of core collapse. The explosion is assumed to release 
a total energy of $E_{\rm ej} = 10^{51}\, \mathrm{erg}$.
When the forward shock of the expanding supernova blast wave reaches a radius of 
$1.2\, \mathrm{pc}$ from the center of the explosion, the simulation transitions 
to a 3D Cartesian coordinate system spanning a domain of $[-50, 50]^3\, \mathrm{pc}$. 
This computational grid consists of $2048^3$ uniformly spaced zones. The initial conditions 
for this phase include the circumstellar medium at the pre-supernova stage, reconstructed 
from the previously described $2.5$D simulation under the assumption of rotational symmetry. 
The angle $\theta_{\rm mag}$, defined as the angle between the symmetry axis of the 
circumstellar medium (aligned with the progenitor’s motion) and the $Oz$ axis of the 
Cartesian grid, is treated as a free parameter in this study.

\subsection{Pulsar wind interaction with the supernova ejecta}
\label{method_psr}

Following the core-collapse supernova explosion, the emergence of the pulsar wind 
is introduced after a delay of approximately $20\, \rm yr$, corresponding to the 
free expansion phase of the remnant. The wind is injected isotropically within a 
spherical region of radius $0.4\,\rm pc$ under non-relativistic conditions, following 
the method of \citet{komissarov_mnras_349_2004}. Although this simplification omits 
relativistic corrections, it provides a numerically stable and physically consistent 
framework for capturing the large-scale interaction between the pulsar wind and 
the surrounding supernova ejecta~\citep{Del_Zanna_etal_2006A&A...453..621D}. 
We adopt a single ideal-gas equation of state throughout the computational domain with 
adiabatic index $\gamma=5/3$, which is a common simplification in global modelling of 
composite supernova remnants, although the material of the pulsar wind is expected to 
behave closer to a relativistic gas of adiabatic index $\gamma=4/3$. 
Using $\gamma=5/3$ instead of $\gamma=4/3$ increases the gas pressure by a factor of 
$\approx 2$, which can mildly bias the effective expansion of the nebula, its morphological 
and emission properties. Our work focuses on the large scale shaping by the pre-supernova 
circumstellar cavity, the motion of the progenitor star and the ISM magnetization, we do not 
expect this approximation to qualitatively modify the trends reported in this work.

The pulsar is initialized with an initial spin period of $P_{\rm o} = 0.3\, \rm s$ 
and a spin-down rate $\dot{P}_{\rm o} = 10^{-17}\, \rm s\, \rm s^{-1}$, under 
the standard assumption of a braking index $n = 3$, indicative of magnetic 
dipole radiation. 
This yields a spin-down luminosity of 
$\dot{E}_{\rm o} = 10^{38}\, \rm erg\, \rm s^{-1}$ at the onset of the pulsar 
wind nebula phase. 
Following \citet{komissarov_mnras_349_2004}, we inject a magnetised wind that 
is initially made of a purely toroidal component with a polar dependence that is 
symmetric in geometry accross the equator. 
The velocity of the injected wind is taken to be $0.01c$, 
where $c$ is the speed of light. Although this is sub-relativistic, it remains 
sufficient to produce the expected expansion behavior of young plerions. 
Such pulsar properties are taken from \citet{2017hsn_book_2159S} and have been 
used in \citet{2025A&A...696L...9M}. 
Hence, this study does not consider yet the pulsar properties as free parameters;
which could potentially affect the results. For instance, a less luminous pulsar 
will inefficiently prevent the reflection of the supernova blastwave and generate, 
at similar time after the explosion, a smaller nebula.

The magnetization parameter is set to $\sigma = 10^{-3}$. This choice prioritizes 
numerical robustness in these global, long-term ($\ge 50$$-$$80\, \rm kyr$) multi-dimensional
runs, and limits sensivity to resolution-dependent in the region near the pulsar wind 
termination shock and in the other strongly 
magnetized shear layers of the supernova remnant. Note that $\sigma \ge 10^{-2}$ is often 
required in axisymmetric (and some 3D) studies to obtain magnetically collimated, hoop-stress-driven 
polar elongations in the inner 
nebula~\citep{del_zanna_421_2004,Bucciantini_aa_422_2004,Porth_etal_2014MNRAS.438..278P}. 
This work does not aim to reproduce that inner jet-launching regime, and, 
the bipolar structures we refer to in the discussion below relate to large scale  
polar outflows guided by the cavity geometry and the total pressure distribution.

We assume a symmetric explosion, such that the neutron star remains fixed at 
the center of the computational domain. Its rotation axis is aligned with the 
$Oz$ direction of the 3D Cartesian grid, which facilitates a clean exploration 
of the impact of the misalignment between the ISM magnetic field and the pulsar’s 
spin axis. This angle, $\theta_{\rm mag}$, is varied in the simulations to 
investigate its effect on the shape, structure, and emission properties of the 
resulting plerion.

\begin{figure*}
        \centering
        \includegraphics[width=0.96\textwidth]{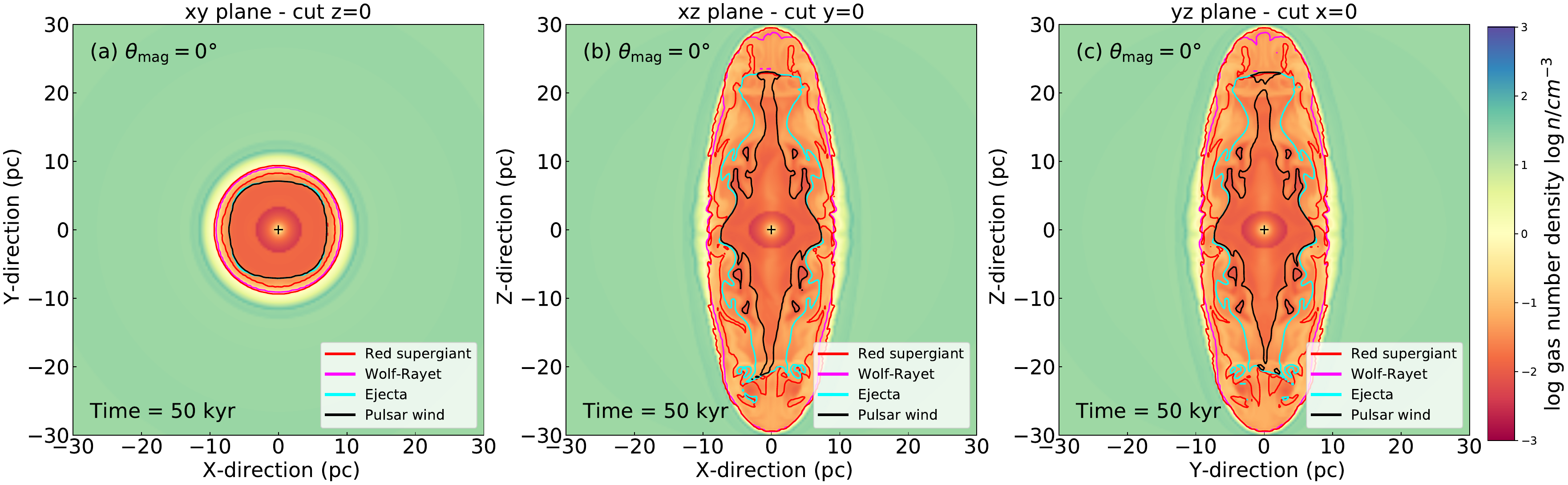}  \\
        \caption{
The number density fields in the 3D supernova remnant models of a static  
$35\, \rm M_{\odot}$ star rotating with $\Omega_{\star}/\Omega_{\rm K} = 0.1$. 
The ISM magnetic field is aligned with the $Oz$ Cartesian axis and has a 
strength of $B_{\rm ISM} = 7\, \rm \mu G$. 
The figures show the supernova remnant in the $z = 0$ (left), 
$y = 0$ (middle), and $x = 0$ (right) planes of the Cartesian coordinate 
system, at $50\, \rm kyr$ after the onset of the explosion. 
The results are shown for an angle $\theta_\mathrm{mag} = 0\degree$ 
between the pulsar's rotation axis and the direction of the local ISM. 
The different contours indicate regions of the supernova remnant with 
a $50\%$ contribution from pulsar wind (black) and ejecta (cyan), 
as well as regions with a $10\%$ contribution from Wolf–Rayet (magenta) 
and red supergiant wind (red) material, respectively. 
The black cross marks the position of the supernova explosion. 
        }
        \label{fig:models_20kyr_density_0deg}  
\end{figure*}

\subsection{Emission maps}
\label{method_emission_maps}

In order to directly compare the numerical models of pulsar wind nebulae developed 
in this study with real astronomical observations, non-thermal emission maps in the 
radio waveband are generated. These maps provide insight into the spatial distribution 
and intensity of synchrotron radiation emitted by non-thermal electrons interacting 
with the magnetic field within the plerion. 
The calculations are carried out using the radiative transfer code 
{\sc RADMC-3D}~\citep{dullemond_2012}, which allows for detailed modeling of 
radiation processes in astrophysical environments. The methodology used to compute the 
synchrotron emission coefficients, which quantify the amount of radiation produced 
per unit volume, follows the approach described in~\citet{meyer_mnras_502_2021}. This 
procedure accounts for both the energy distribution of relativistic electrons 
and the local magnetic field strength and structure. The method is based on the 
formalism outlined in~\citet{villagran_mnras_527_2024} and~\citet{meyer_2024_loops}.
To produce the intensity maps, the computed emissivity distribution is projected 
onto the plane of the sky, considering the inclination angle of the supernova remnant 
relative to the observer’s line of sight. 
Because the dynamics are computed in the non-relativistic magneto-hydrodynamics 
approximation and we do not apply the Doppler boosting or any aberration corrections 
in the post-processing, the synthetic maps are intended for qualitative discussion 
of the large-scale morphology at this stage, rather than for a detailed test of the asymmetries of 
the brightnesses of the inner nebula.

\subsection{Numerical methods}
\label{method_numerical}

To ensure the fidelity of the interaction between the pulsar wind and the interstellar 
medium when simulating the evolution of the pulsar wind nebula, the computational grid 
for the circumstellar medium is assigned a spatial resolution exceeding that of the 
plerionic supernova remnant. This refinement guarantees that the termination shock 
of the pulsar's wind nebula is sufficiently resolved, mitigating numerical boundary 
effects that could otherwise introduce artificial structures.  
The numerical framework is constructed upon a dimensionally unsplit, finite-volume 
solver based on the Godunov method. The system integrates the Harten–Lax–van 
Leer approximate Riemann solver \citep{hll_ref} for flux computations, coupled with a 
third-order Runge–Kutta temporal integration scheme. Reconstruction of primitive variables 
between adjacent computational cells follows a hybrid methodology: the circumstellar medium 
employs a piecewise parabolic method to enhance accuracy, whereas the supernova remnant 
is computed using a linear reconstruction scheme constrained by a minmod slope limiter. 
The latter approach enforces a total variation diminishing condition, preserving numerical 
stability in regions exhibiting steep gradients.  
Temporal evolution adheres to the Courant–Friedrichs–Levy condition, which is initially 
defined by a number of 0.3 to maintain stability while maximizing computational 
efficiency. Furthermore, divergence-free preservation of the magnetic field vector 
throughout the domain is enforced through the divergence-cleaning methodology of 
Powell \citep{Powell1997}, ensuring compliance with the solenoidal constraint 
in magnetohydrodynamic simulations.


\section{Results}
\label{results}

\subsection{Effects of the magnetic field direction on the properties of the supernova remnants}
\label{resultats_effects}

Fig.~\ref{fig:models_20kyr_density_0deg} shows the number density fields in 3D plerionic 
supernova remnant models of a static $35\, \rm M_{\odot}$ star rotating at an equatorial 
rate of $\Omega_{\star}/\Omega_{\rm K} = 0.1$.  
This baseline model represents a supernova remnant and its pulsar wind nebula evolving in 
an ISM with a magnetic field of strength $B_{\rm ISM} = 7\, \rm \mu G$, aligned with the 
$Oz$ axis of the Cartesian coordinate system—which, in this case, also corresponds to the 
rotation axis of the pulsar.  
The panels display slices of the number density field in the 
$z = 0$ (Fig.~\ref{fig:models_20kyr_density_0deg}a), 
$y = 0$ (Fig.~\ref{fig:models_20kyr_density_0deg}b), and 
$x = 0$ (Fig.~\ref{fig:models_20kyr_density_0deg}c) planes, 
taken at $50\, \rm kyr$ after the onset of the supernova explosion.  
The different contours indicate regions of the supernova remnant with a $50\%$ number 
density contribution from the pulsar wind (black) and the supernova ejecta (cyan), as well 
as a $10\%$ contribution from Wolf–Rayet (magenta) and red supergiant wind (red) material.  
The cavity carved out during the progenitor’s lifetime is bounded by the termination shock 
of the stellar wind bubble, which is elongated along the direction of the ISM magnetic field  \citep{vanmarle_584_aa_2015}. This cavity appears tubular due to the 2.5-dimensional nature 
of the circumstellar medium simulation \citep{meyer_2024_loops} 
(see Fig.~\ref{fig:models_20kyr_density_0deg}a), allowing the formation of a pulsar wind 
nebula composed of an equatorial disc and a \textcolor{black}{polar elongation} extending up to $\pm 25\, \rm pc$ 
along the $Oz$ direction, rather than exhibiting the diamond-like morphology described by 
\citet{komissarov_mnras_349_2004}.

Figs. \ref{fig:models_20kyr_density_30deg} and \ref{fig:models_20kyr_density_45deg} are similar to 
Fig. \ref{fig:models_20kyr_density_0deg}, with angles $\theta_\mathrm{mag} = 30\degree$ and 
$\theta_\mathrm{mag} = 45\degree$ between the local ISM field and the pulsar rotation axis, 
respectively. The intersections of the tubular cavity of stellar wind, in which the supernova 
ejecta and the pulsar wind were released and evolved, with the $z = 0$ and $y = 0$ planes are 
ellipses (Figs. \ref{fig:models_20kyr_density_30deg}a,b and \ref{fig:models_20kyr_density_45deg}a,b). 
In contrast, the intersection with the $x = 0$ plane corresponds to the same tubular cavity as 
in Fig. \ref{fig:models_20kyr_density_0deg}b,c, rotated by angles $\theta_\mathrm{mag} = 30\degree$ 
and $\theta_\mathrm{mag} = 45\degree$, respectively.
The morphology of the pulsar wind nebula is strongly influenced by the direction of the cavity in 
which the material can freely expand. Specifically, the disc no longer lies in the equatorial plane 
of the rotating pulsar; instead, it grows in a plane that forms an angle $\theta_\mathrm{mag}$ with 
the $z = 0$ plane. With more space available, as it is no longer confined by the walls of the stellar 
wind cavity (see Fig. \ref{fig:models_20kyr_density_0deg}b,c), the pulsar wind nebula disc is larger 
and extends to distances twice that of the model with $\theta_\mathrm{mag} = 0\degree$.
Similarly, the \textcolor{black}{polar elongations} of the pulsar wind are affected by the distribution of circumstellar 
material and bend due to the ram pressure exerted by the defunct stellar wind and the supernova 
ejecta reflecting within the cavity. This pressure compresses the \textcolor{black}{polar elongation} and channels it towards the 
end of the cavity, naturally producing a complex shape with a wobbled disc and a bent bipolar 
structure. The outer extent of the bipolar structure realigns with the direction of the local 
ISM magnetic field (see Fig. \ref{fig:models_20kyr_density_30deg}c).

Finally, Fig. \ref{fig:models_20kyr_density_90deg} presents the pulsar wind nebula modeled with 
$\theta_\mathrm{mag} = 90\degree$. The intersection of the circumstellar medium with the $x = 0$ 
plane appears as a circle and shares the same elongated cavity shape as shown in 
Fig. \ref{fig:models_20kyr_density_0deg} for the other planes.
The morphology of the pulsar wind nebula exhibits an equatorial disc that is significantly more extended 
than the \textcolor{black}{polar elongation} originating from it (Fig. \ref{fig:models_20kyr_density_90deg}a). This configuration 
results from the orientation of the circumstellar medium, which is influenced by the direction of 
the local magnetization of the ISM. The diamond-shaped structure identified in the study by 
\citet{komissarov_mnras_349_2004} reappears; however, the two components of the \textcolor{black}{polar elongation} are thicker 
and less affected by instabilities. Additionally, the space occupied by the disc adopts a rectangular 
morphology with protuberances along the $y$-axis direction \citep{meyer_527_mnras_2024}, as 
depicted in Fig. \ref{fig:models_20kyr_density_90deg}ac.

\begin{figure*}
        \centering
        \includegraphics[width=0.96\textwidth]{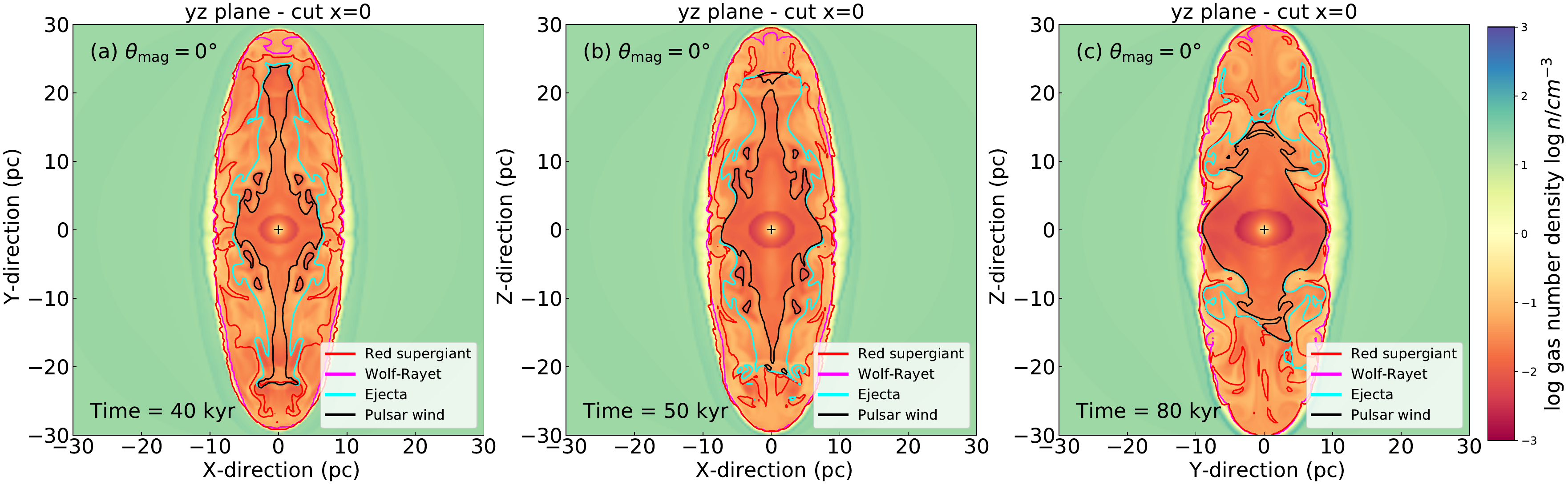}  \\        
        \caption{
Time evolution of the number density fields in the 3D supernova remnant models of a static  
$35\, \rm M_{\odot}$ star rotating with $\Omega_{\star}/\Omega_{\rm K} = 0.1$. 
The ISM magnetic field is aligned with the $Oz$ Cartesian axis and has a 
strength of $B_{\rm ISM} = 7\, \rm \mu G$.
The figures show the supernova remnant in the $z = 0$ (left), 
$y = 0$ (middle), and $x = 0$ (right) planes of the Cartesian coordinate 
system, spanning times from $40\, \rm kyr$ to $80\, \rm kyr$ 
after the onset of the explosion. 
The results are shown for an angle $\theta_\mathrm{mag} = 0\degree$ 
between the pulsar's rotation axis and the direction of the local ISM. 
The different contours indicate regions of the supernova remnant with 
a $50\%$ contribution from pulsar wind (black), ejecta (cyan), 
Wolf–Rayet (magenta), and red supergiant wind (red) material, respectively. 
The black cross marks the position of the supernova explosion. 
        }
        \label{fig:models_0deg_time_evolution}  
\end{figure*}

\begin{figure*}
        \centering
        \includegraphics[width=0.49\textwidth]{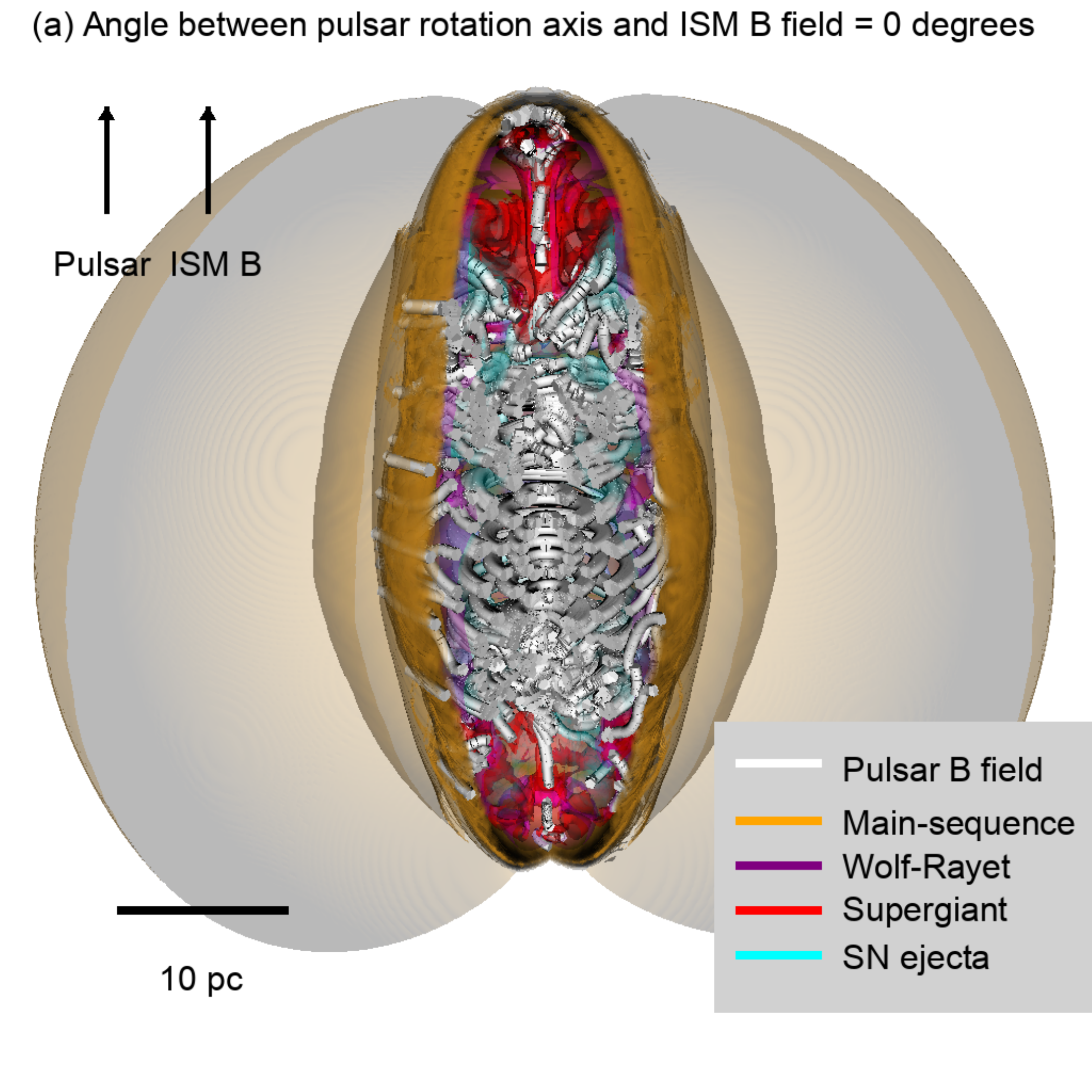} 
        \includegraphics[width=0.49\textwidth]{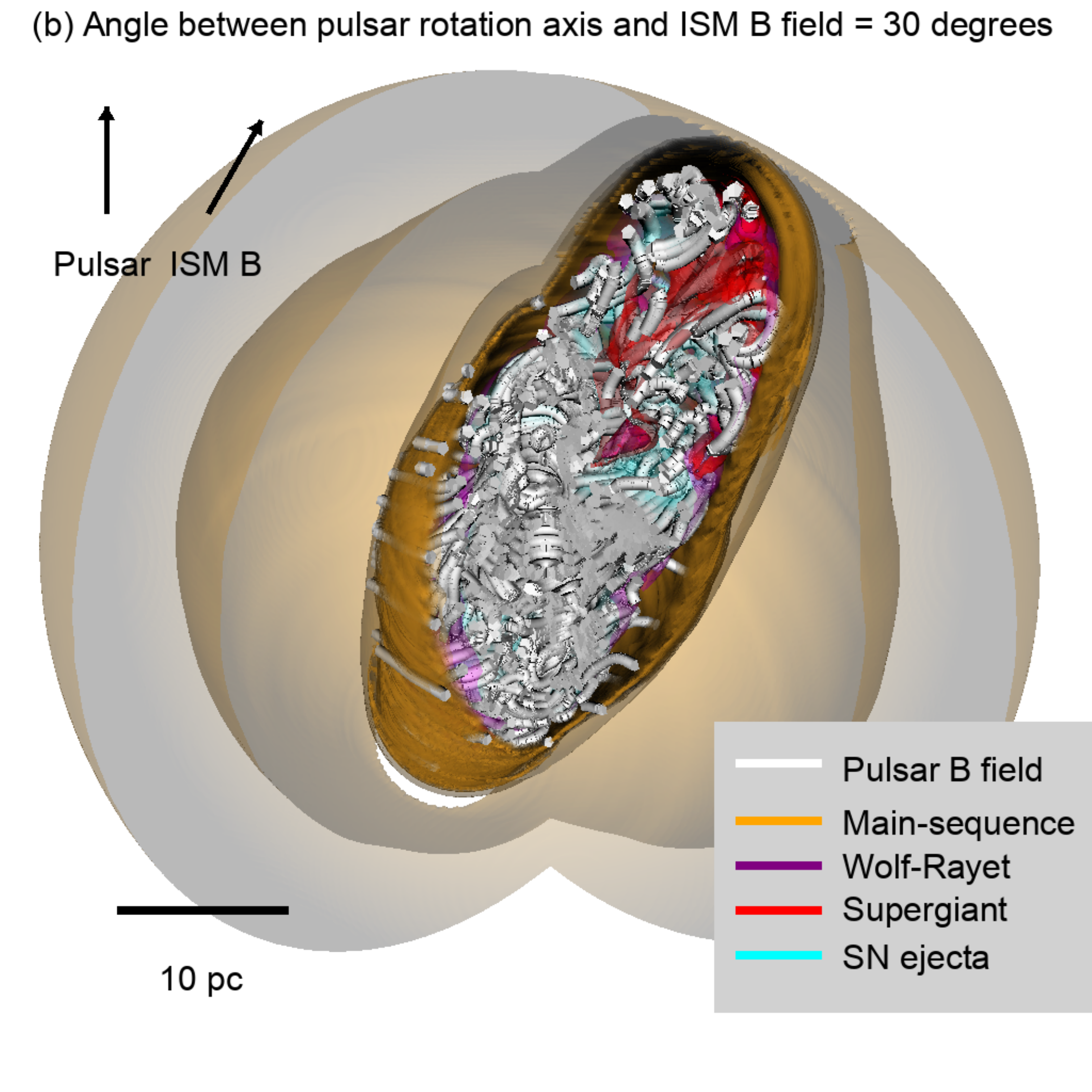}  \\
        \includegraphics[width=0.49\textwidth]{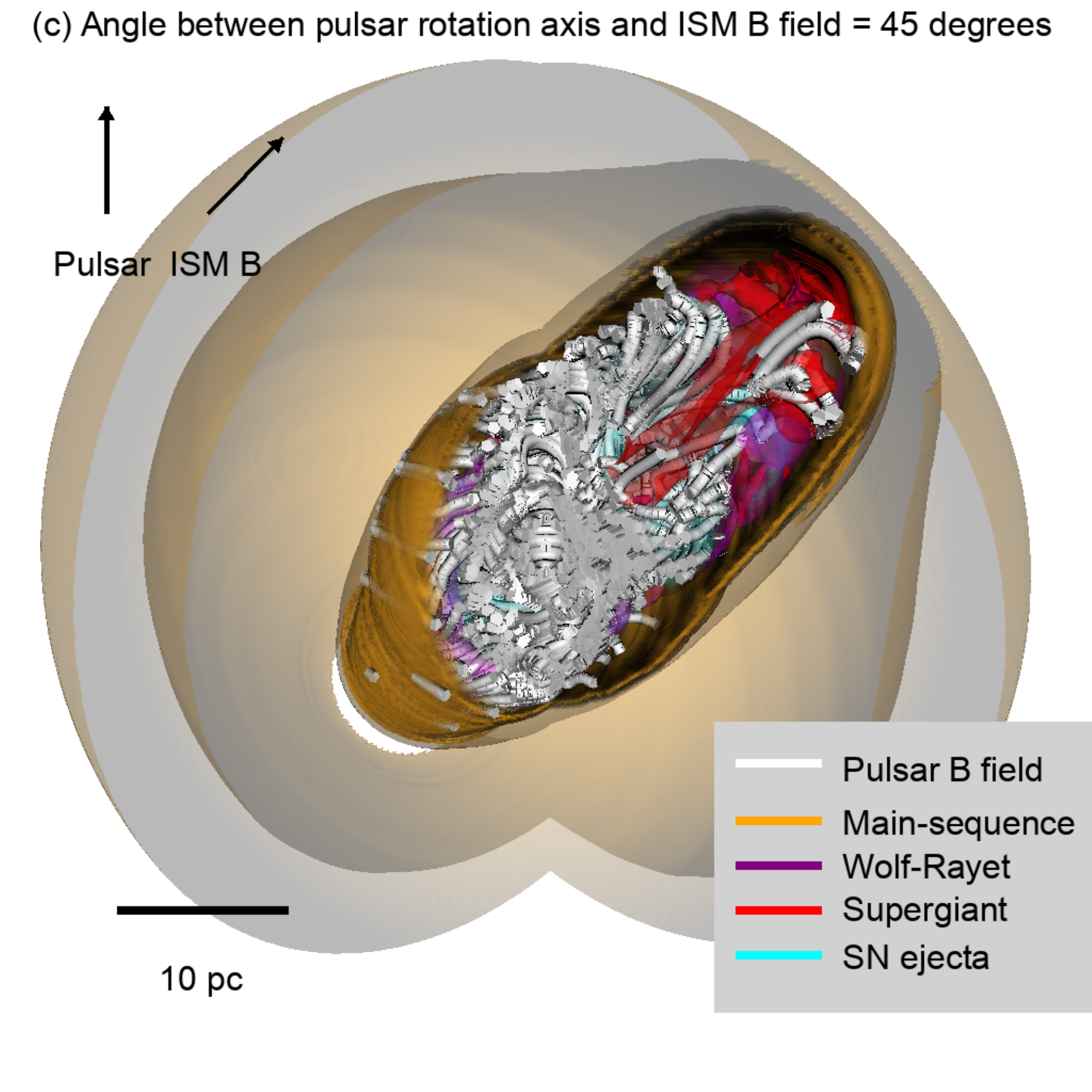}  
        \includegraphics[width=0.49\textwidth]{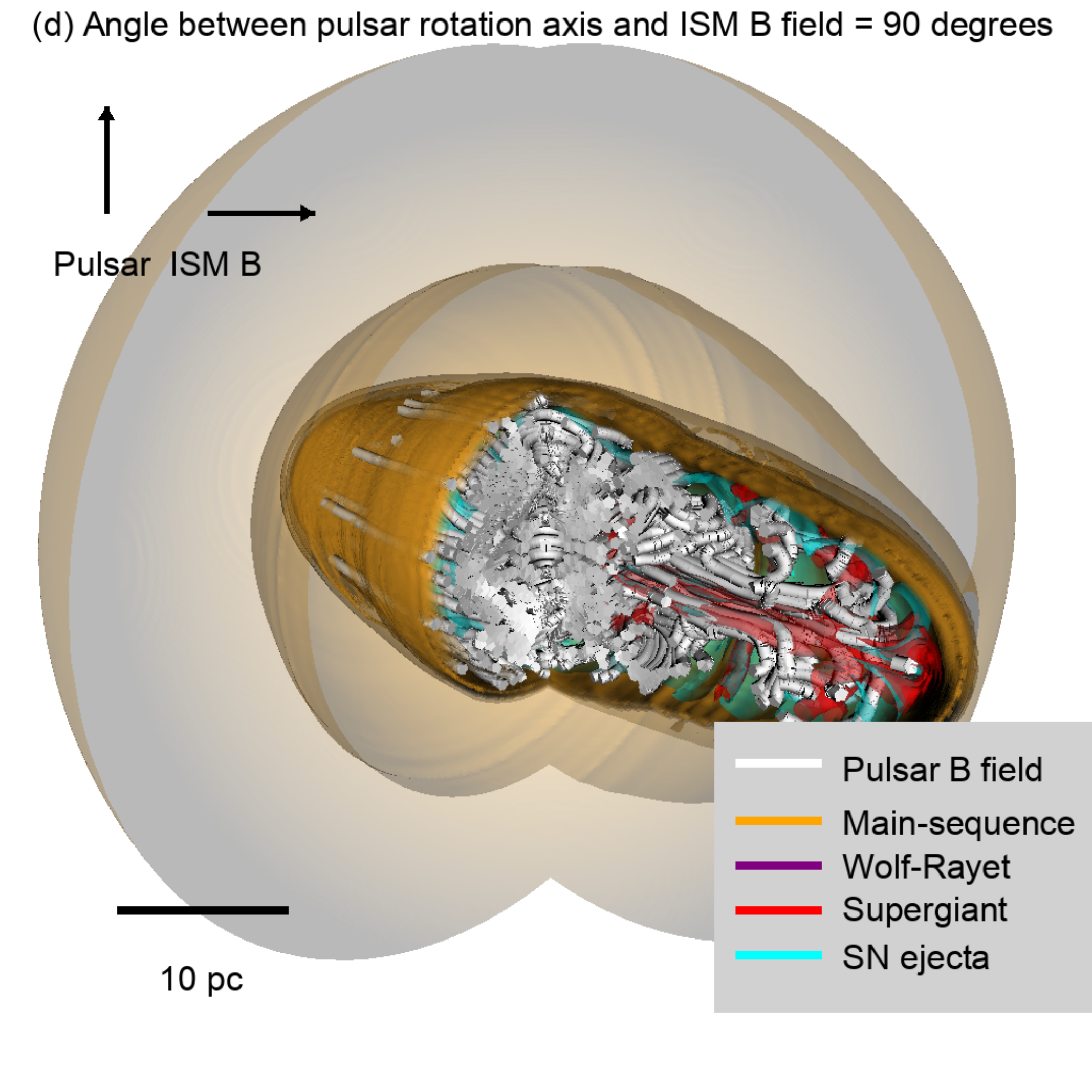} \\ 
        \caption{
        Rendering of the 3D MHD pulsar wind nebulae and their complex environments. 
        The orange surfaces trace constant density regions of the stellar wind bubble, 
        the red surfaces trace the red supergiant wind, the magenta surface trace the Wolf-Rayet 
        wind, the cyan surface traces the supernova ejecta and the white tubes trace the 
        magnetic field in the pulsar wind. The 3D figures are clipped with two planes 
        permitting a visualization of the internal structure of the plerionic supernova remnant. 
        The displayed models differ only by the angle between the ISM magnetic field lines 
        that shape the main-sequence stellar wind bubble (orange) and the axis of rotation 
        of the pulsar (fixed in all models as being along the vertical direction). 
        The angle is of $\theta_\mathrm{mag}=0\degree$ (a), 
        $\theta_\mathrm{mag}=30\degree$ (b), 
        $\theta_\mathrm{mag}=45\degree$ (c) 
        and $\theta_\mathrm{mag}=90\degree$ (d), respectively. 
        }
        \label{fig:3D_PWN_rendering}  
\end{figure*}

\begin{figure}
        \centering
        \includegraphics[width=0.4\textwidth]{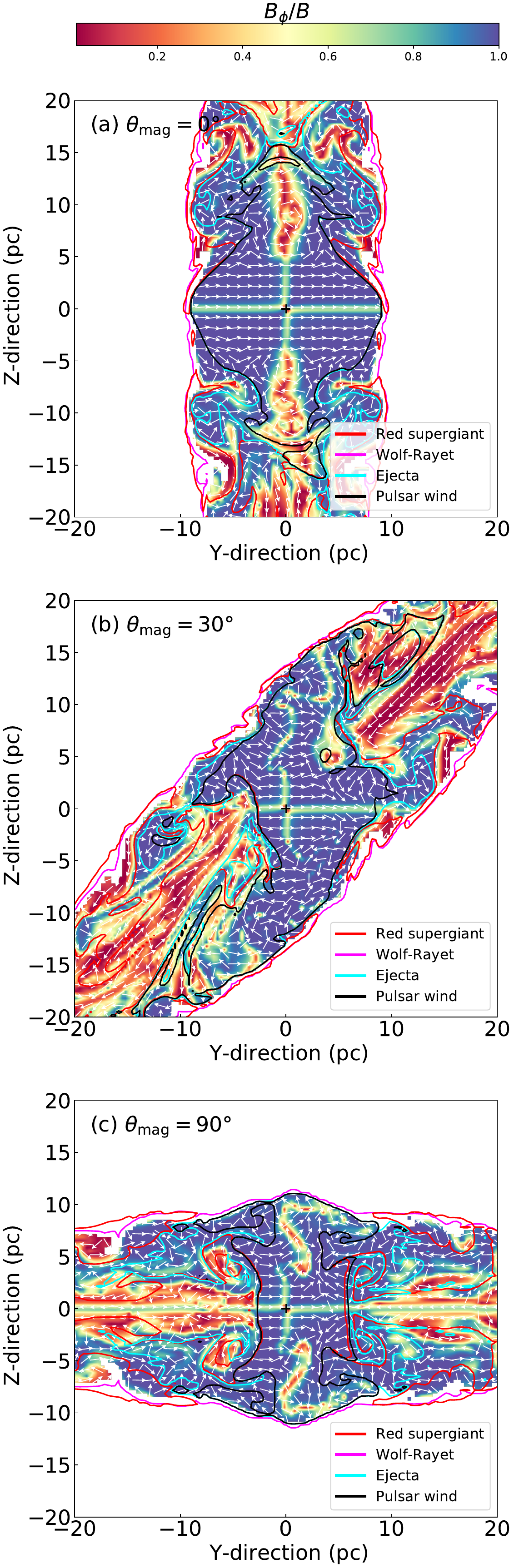}  \\        
        \caption{
Distribution of the toroidal to total magnetic field ratio $B_{\phi}/B$ in the 
3D supernova remnant models. The white arrows trace the magnetic field in the yOz 
plane for the region of the supernova remnant. 
The figures show the supernova remnants in the $x = 0$ plane of the Cartesian 
coordinate system, at time $70\, \rm kyr$ after the onset of 
the explosion. 
        }
        \label{fig:magnetic}  
\end{figure}

\begin{figure*}
        \centering
        \includegraphics[width=0.49\textwidth]{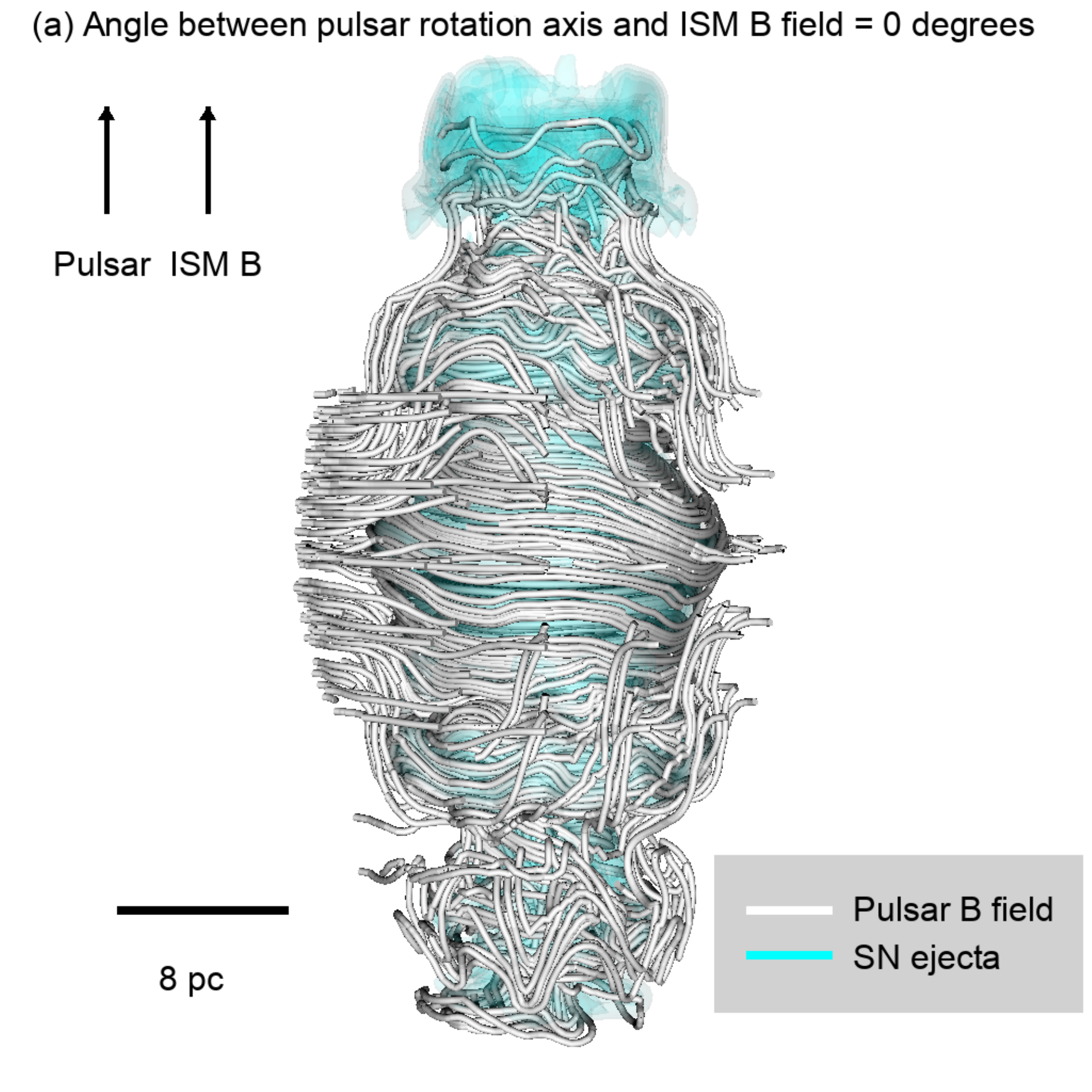}  
        \includegraphics[width=0.49\textwidth]{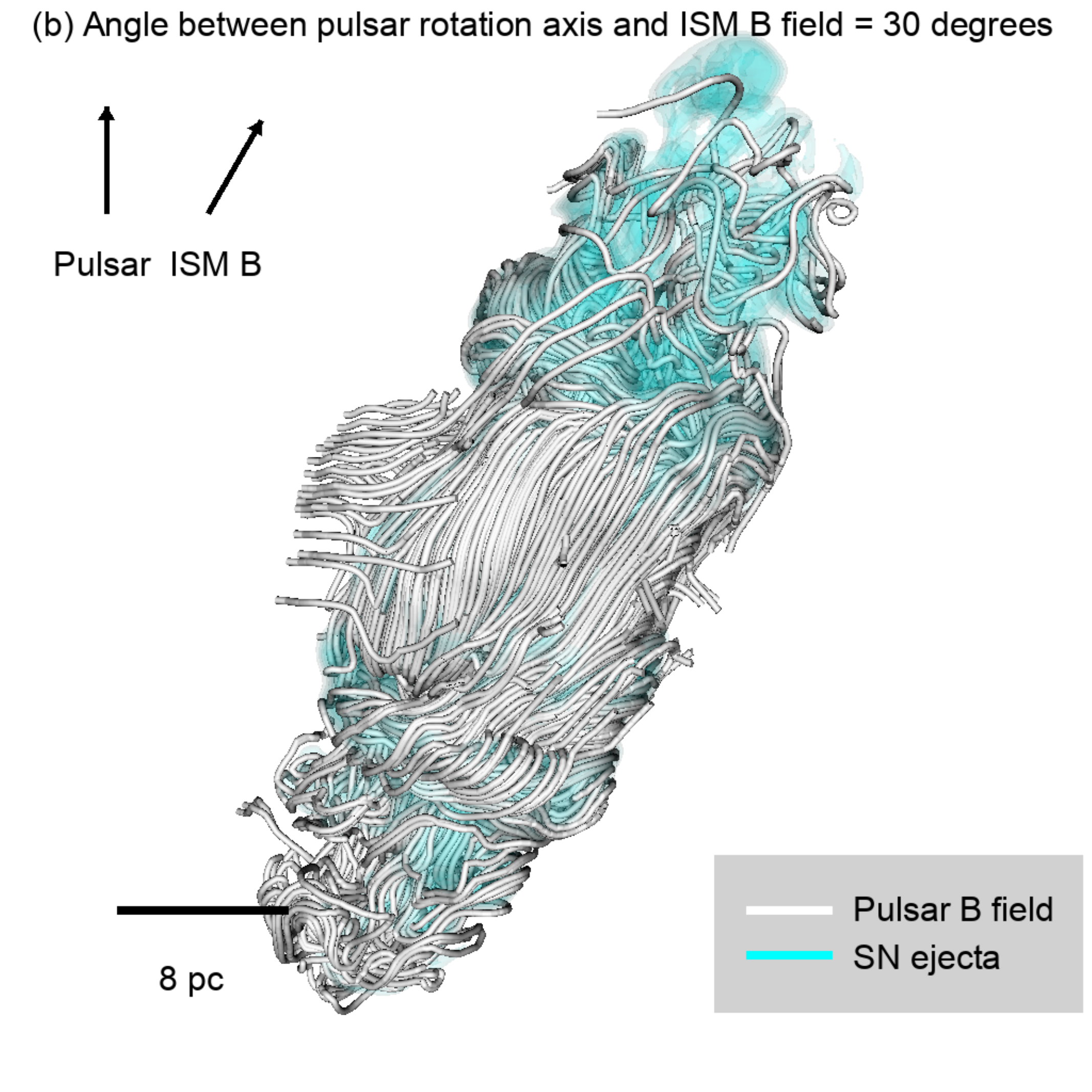}  \\
        \includegraphics[width=0.49\textwidth]{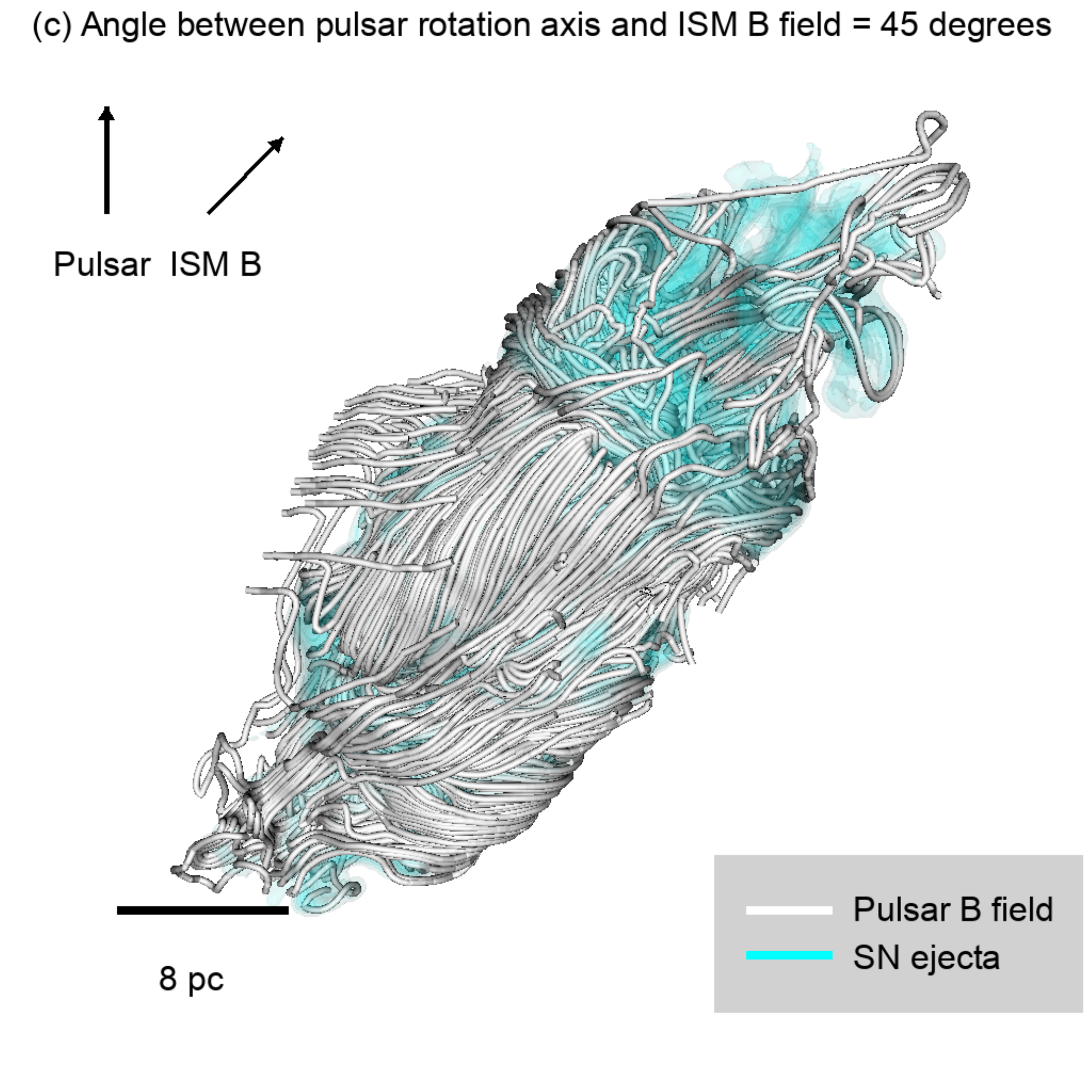}  
        \includegraphics[width=0.49\textwidth]{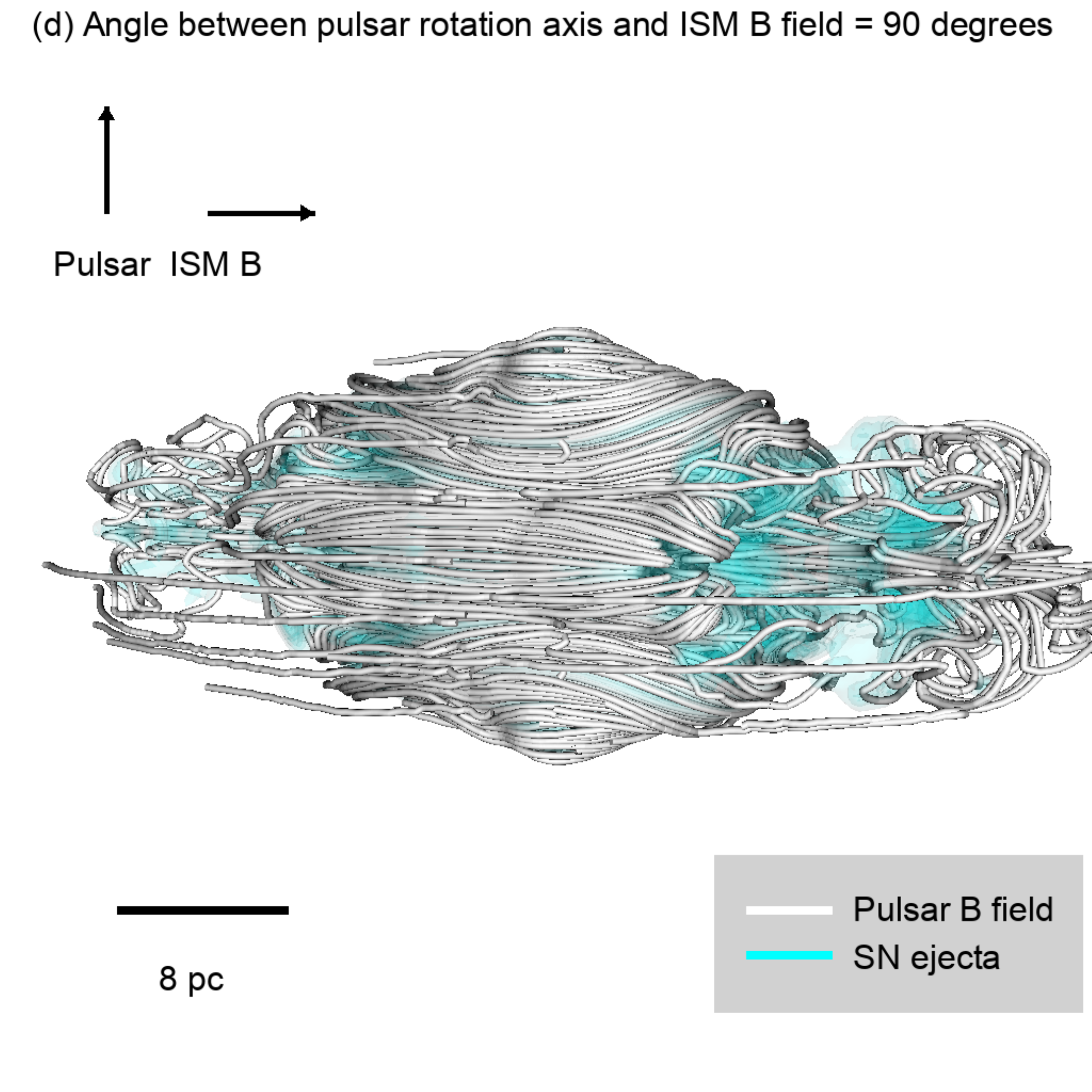} \\ 
        \caption{
        Same as Fig. \ref{fig:3D_PWN_rendering}, displaying the 3D magnetic 
        field distribution of the the pulsar wind nebulae within the supernova ejecta.  
        It highlights how the supernova ejecta distribution (itself governed by the 
        circumstellar medium and the magnetization of the ISM) influences the 
        manner the propagation of the pulsar wind and the arrangement of its magnetic 
        field lines. 
        An animated version of this figure is available as online material. 
        }  
        \label{fig:3D_PWN_B_field}  
\end{figure*}

\begin{figure*}
        \centering
        \includegraphics[width=0.95\textwidth]{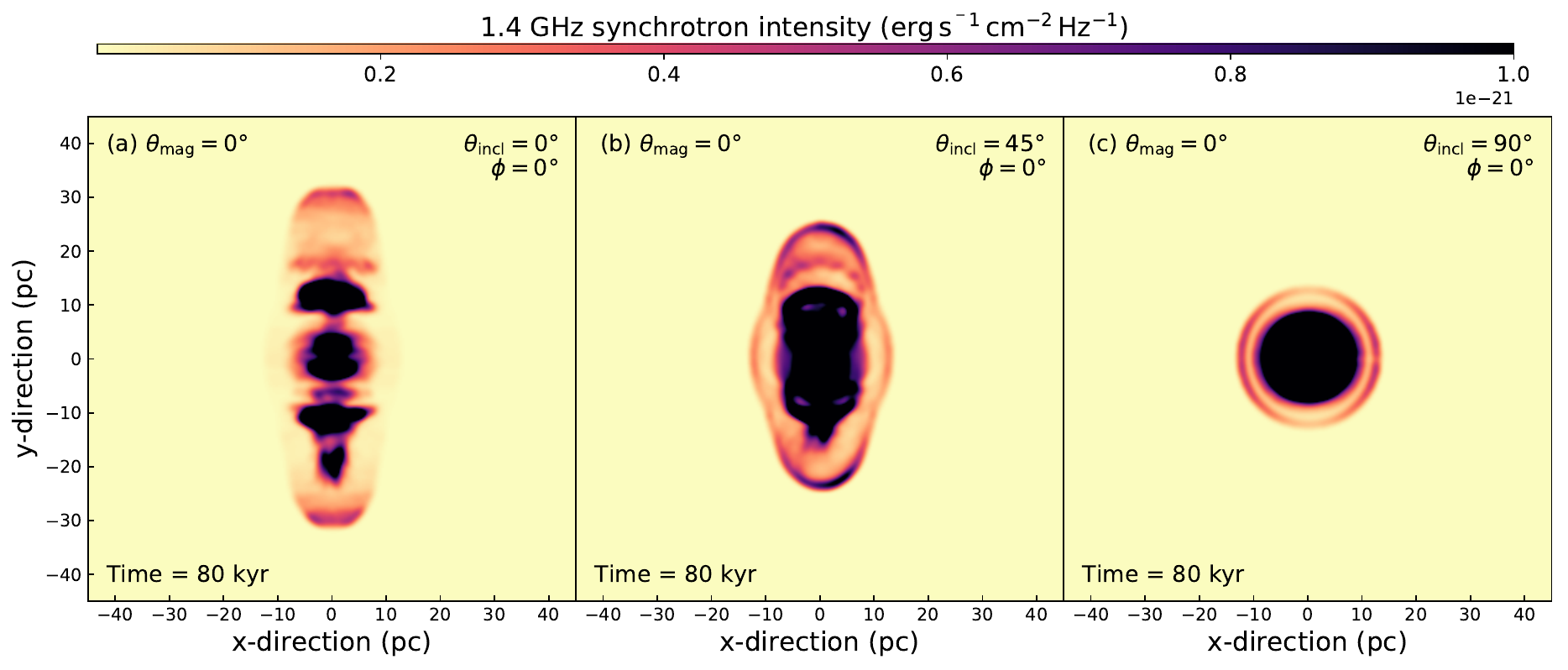}  \\
        \caption{
        Synchrotron $1.4\, \rm GHz$ radio emission maps of the model with $\theta_\mathrm{mag}=0\degree$ 
        at times $80\, \rm kyr$ as seen with an angle between the observers point of view and the 
        plan in which the supernova remnant is projected $\theta_\mathrm{incl}=0\degree$.   
        }
        \label{fig:emission_maps_0deg}  
\end{figure*}

\subsection{Time evolution of the plerionic supernova remnant}
\label{resultats_evolution}

Fig. \ref{fig:models_0deg_time_evolution} shows the time evolution of the pulsar wind nebula with
$\theta_\mathrm{mag}=0\degree$, from $40, \rm kyr$ (Fig. \ref{fig:models_0deg_time_evolution}a) to
$80, \rm kyr$ (Fig. \ref{fig:models_0deg_time_evolution}c) after the supernova explosion, as shown
in the $x=0$ plane. The disc of the pulsar wind nebula grows along the $Ox$ direction.
It has been demonstrated that the supernova shock wave reflects off the walls of the stellar wind 
cavity, and that a wave propagates back to the explosion center, inducing the formation of rectangular 
supernova remnants \citep{2022MNRAS.515..594M}. Additionally, the release of a pulsar wind after the 
explosion further shapes the plerion \citep{meyer_527_mnras_2024}. In this scenario, the pulsar wind, 
initially stretched along $Oy$ but relatively compact along $Ox$, interacts with the reflected waves 
of the forward shock that has impacted the termination shock of the supernova blastwave 
(Fig. \ref{fig:models_0deg_time_evolution}a,b).
A similar phenomenon occurs with the \textcolor{black}{polar elongation} interacting with the end of the circumstellar cavity 
previously filled with ejecta. This interaction leads to the detachment of clumps rich in pulsar wind 
material, resulting in a smaller, denser pulsar wind nebula. Consequently, the disc-to-\textcolor{black}{polar elongation} size 
ratio becomes larger compared to earlier evolutionary phases (Fig. \ref{fig:models_0deg_time_evolution}c).

Fig. \ref{fig:models_30deg_time_evolution}a depicts the simulation with $\theta_\mathrm{mag}=30\degree$, 
where the misalignment between the pulsar's equatorial plane and either the equatorial plane of the 
pulsar wind nebula or the direction of the \textcolor{black}{polar elongations}—both channeled through the low-density region 
of stellar wind—is evident at $40, \rm kyr$ post-explosion, persisting thereafter 
(Fig. \ref{fig:models_30deg_time_evolution}b,c).
As time progresses, the pulsar wind nebula's morphology becomes increasingly distorted, adopting the 
shape of a complex region with a squared core and arms emerging from its edges, particularly distorted 
at the bottom of the circumstellar cavity by $80, \rm kyr$ (Fig. \ref{fig:models_30deg_time_evolution}c).
In the model with $\theta_\mathrm{mag}=45\degree$, the angle between the pulsar's rotation axis and 
the local interstellar medium (ISM) magnetic field direction reduces the alignment between the disc 
and the \textcolor{black}{polar elongation}. At $40, \rm kyr$, the \textcolor{black}{polar elongation} splits into two smaller filaments 
(Fig. \ref{fig:models_45deg_time_evolution}a), which merge into a single structure by 
$50, \rm kyr$ (Fig. \ref{fig:models_45deg_time_evolution}b). By $80 \rm kyr$, this evolves 
into a central squared structure with two thick \textcolor{black}{polar elongation}s attached, forming a complex pulsar wind 
nebula (Fig. \ref{fig:models_45deg_time_evolution}c).

Fig. \ref{fig:models_90deg_time_evolution}a illustrates the time evolution of the pulsar wind 
nebula for $\theta_\mathrm{mag}=90\degree$. At $40, \rm kyr$, the disc aligns with the pulsar's 
rotation plane, extending farther than the \textcolor{black}{polar elongation}. Here, the equatorial plane interacts with 
the end of the circumstellar cavity, becoming unstable and diminishing in size due to reflections 
from both the pulsar wind's forward shock and the supernova blastwave's forward shock.
The \textcolor{black}{polar elongation} evolves as it first impacts the cavity, enlarging the filaments perpendicular to the 
equatorial disc by $50, \rm kyr$ (Fig. \ref{fig:models_90deg_time_evolution}b). The system continues 
to evolve, adopting a squared shape with several \textcolor{black}{polar elongation}s and filaments emerging by $80, \rm kyr$ 
(Fig. \ref{fig:models_90deg_time_evolution}c).

\subsection{Structure of the pulsar magnetic field}
\label{resultats_supernova_remnants}

Fig. \ref{fig:3D_PWN_rendering} displays volumetric renderings of the 3D MHD models for 
pulsar wind nebulae and their complex environments. The colour coding is as follows. 
The orange surfaces mark regions of constant density stellar wind material and therefore 
highlight the location of the stellar wind bubble shaped by the progenitor prior to the explosion. 
The red surface trace the red supergiant wind in $10 \%$ in number density, the magenta 
surface traces the Wolf-Rayet in $10 \%$ in number density, the cyan surface traces the 
the locations of the supernova ejecta being $50 \%$ in number density and the white tubes 
trace the magnetic field lines of the pulsar wind. 
The displayed models differ by the angle between the ISM magnetic field lines shaping the 
main-sequence stellar wind bubble (orange) and the axis of rotation of the pulsar 
(fixed in all models as being along the vertical direction). The angle is of 
$\theta_\mathrm{mag}=0\degree$ (a), $\theta_\mathrm{mag}=30\degree$ (b), 
$\theta_\mathrm{mag}=45\degree$ (c) and $\theta_\mathrm{mag}=90\degree$ (d), respectively. 
Fig. \ref{fig:magnetic} additionally shows the distribution of the toroidal to total 
magnetic field ratio $B_{\phi}/B$, with arrows tracing the magnetic field vectors in the yOz 
plane for the regions of the supernova remnant and contours highlighting the contact 
discontinuities of the several materials mixing into the remnant.

In Figure \ref{fig:3D_PWN_rendering}a, with $\theta_\mathrm{mag}=0^\circ$, the large-scale 
circumstellar wind bubble extends over a spherical region with a radius of approximately 
$30\, \rm pc$. The density surface highlights the outer forward shock of the wind bubble, 
which is axisymmetric due to its pre-calculation in 2.5 dimensions. This configuration results 
in an increase in density, forming oblate shells stretched along the pulsar's rotation axis, 
aligning with the local ISM magnetic field direction.
The concentrations of isodensity surfaces denote the region of the termination shock of the 
stellar wind cavity, which has been impacted by the forward shock of the supernova blastwave. 
Inside this cavity, the red supergiant material has dispersed throughout the available space, 
propelled by the last pre-supernova stellar wind expelled during the Wolf-Rayet evolutionary 
phase. The supernova ejecta fill the entire cavity, mixing with other materials.
The distribution of magnetic field lines exhibits the typical morphology of a toroidal field, 
wrapping around the central pulsar and being displaced by the its wind. 
The region of unmixed pulsar wind (black contour) is dominated by a toroidal field, except 
at the apex of its bipolar elongations. In the extends of the cavity, where ejecta and stellar 
wind meet, the magnetic field is disorganised (Fig. \ref{fig:magnetic}a).  
The pulsar wind predominantly occupies the central spherical region where the supernova 
blastwave first interacted with the stellar wind cavity.

Figs. \ref{fig:3D_PWN_rendering}b,c show the plerionic supernova remnant for 
$\theta_\mathrm{mag}=30^\circ$ and $\theta_\mathrm{mag}=45^\circ$, respectively, where 
the location of the cavity is no longer aligned with the pulsar rotation axis. This has the 
following effect: the distribution of the magnetic field lines loses symmetry but propagates 
towards the end of the cavity, where they reconnect, forming a complex interlace that strongly 
deviates from the situation with $\theta_\mathrm{mag}=0^\circ$ (Fig. \ref{fig:3D_PWN_rendering}a). 
The same occurs in the model with $\theta_\mathrm{mag}=45^\circ$ (Fig. \ref{fig:3D_PWN_rendering}c). 
Fig. \ref{fig:3D_PWN_rendering}d illustrates the simulation with $\theta_\mathrm{mag}=90^\circ$ 
in which the pulsar rotation axis is normal to the direction of the local magnetic field. 
The toroidal component of the pulsar wind magnetosphere extends deep into the cavity, whereas 
the \textcolor{black}{polar elongation} hits its walls and thickens the polar region towards the disc.
The bouncing mixed material of ejecta and stellar wind can is locally organised towards the center of 
the stretched cavity, although mostly made of a poloidal magnetic field (Fig. \ref{fig:magnetic}b).

Fig. \ref{fig:3D_PWN_B_field} is similar to Fig. \ref{fig:3D_PWN_rendering}, showing the volumetric 
rendering of the supernova ejecta (cyan surfaces) and the pulsar wind magnetic field lines (white tubes). 
The toroidal structure of the magnetic fields wraps around the contact discontinuity between 
the supernova ejecta and pulsar wind material, extending up to the end of the stellar wind 
cavity, which is normal to the pulsar equatorial plane, as shown in Fig. \ref{fig:3D_PWN_B_field}a. 
When an angle $\theta_\mathrm{mag}=30^\circ$ is introduced between the direction of the pulsar 
rotation axis and the local ISM magnetic field direction, the distribution of the pulsar wind 
magnetic field lines changes accordingly. These field lines wrap around the contact discontinuity of the 
supernova ejecta, while maintaining their local toroidal direction, reconnecting with the ISM 
magnetic field that has penetrated the post-shock region of the circumstellar wind bubble and 
borders the edge of the stellar wind cavity. 
The field lines are particularly disorganized at the end of the cavity, where the region is more 
turbulent and hosts a lot of mixing of materials 
(Fig. \ref{fig:3D_PWN_B_field}b). A similar situation occurs in the model with 
$\theta_\mathrm{mag}=45^\circ$ (Fig. \ref{fig:3D_PWN_B_field}c). 
The situation with $\theta_\mathrm{mag}=90^\circ$ resembles that with $\theta_\mathrm{mag}=0^\circ$ 
except that the toroidal field lines wrap around the entire contact discontinuity of the supernova 
ejecta, which is aligned with the pulsar equatorial plane (Fig. \ref{fig:3D_PWN_B_field}d).
Both the vicinity of the pulsar as well the regions close to the discontinuities where a lot of 
mixing happens are filled with toroidaly magnetised material, while the bouncing supernova ejecta 
are mostly poloidaly magnetised (Fig. \ref{fig:magnetic}d).


\section{Discussion}
\label{discussion}

\subsection{Limitations of the models}
\label{limitations}

The limitations of the simulations presented in this paper are threefold. First, the axisymmetric 
nature of the circumstellar medium used as initial conditions for the pulsar wind nebula model; 
second, the incomplete treatment of microphysical processes included in the simulations; and third, 
the static nature of the pulsar.  Let us discuss these points in more detail.

The 2.5D magnetohydrodynamical simulations of the circumstellar medium require models in which 
the motion of the progenitor star aligns with the direction of the ISM magnetic field. A fully 
3D representation of the stellar environment during the pre-supernova phase~\citep{geen_mnras_448_2015} 
would overcome this constraint and improve the accuracy of the modeled plerions, particularly 
regarding the interaction of the supernova blast wave with the walls of the stellar cavity. 
A desired improvement, today numerically unreachable, would be to consider this problem 
in the frame of relativistic magneto-hydrodynamics, adressing amongst others, the 
effects of the adiabiatic index on the structure and emission properties of plerionic 
supernova remnants. 
In addition, relativistic effects, principally relevant for the emission of the inner 
pulsar wind nebula such as the Doppler boosting or emission aberration are not included yet 
in the construction of the synchrotron maps. This can alter the apparent 
distribution of the projected emission in the \textcolor{black}{polar elongation}/torus regions and therefore limits 
direct one-to-one comparison to real high-resolution pulsar wind nebula images, which 
are out of the scope of this study.

Although our models incorporate magnetic fields along with optically thin radiative cooling 
and heating, other physical mechanisms, such as electron heat conduction or non-ideal 
magnetohydrodynamical effects, are neglected and should ideally be considered in future studies.  
Thermal conduction will smooth discontinuities and transitions in the density and the temperature 
fields, in particular at the termination shocks and inside of the hot layers of low-density 
material~\citep{balsara_mnras_386_2008,meyer_2014bb}, the latter will add realism to the 
local magnetic field properties and its effects on the growth of dust 
grains~\citep{gutierrez_aa_670_2023,giang_mnras_530_2024}.

The pulsar considered in our simulations is static and does not move relative to the progenitor 
star; that is, the natal kick imparted during the supernova explosion is not accounted for. 
However, pulsars typically receive kick velocities of several hundred 
$\rm m\,s^{-1}$~\citep{verbunt_aa_608_2017}, which can displace them by up to $10\, \rm pc$ 
within the supernova remnant~\citep{igoshev_mnras_494_2020,igoshev_mnras_494_2020b}. 
Given the size of the cavity carved by the progenitor in the ISM, a moving pulsar would 
still remain within it, and would thus have only a second-order effect on the overall 
morphology of the pulsar wind nebula~\citep{meyer_527_mnras_2024}. 
The star itself, but also the local inhomogeneities of the ISM, also influence the 
emission of the pulsar wind nebula, via their control onto the circumstellar 
distribution of material at the pre-supernova time. They constitute a vast 
parameter space needs to be explored. 
All these factors (Fig. \ref{fig:concept_plot_caveats}) could enhance the realism 
of the simulations, and, in turn, improve the accuracy of the resulting synthetic 
emission maps when compared with real observations.

\subsection{Synchrotron maps}   
\label{resultats_maps}

Fig.~\ref{fig:emission_maps_0deg} displays synchrotron $1.4\, \rm GHz$ radio emission maps 
of the model with $\theta_\mathrm{mag}=0\degree$ at $80\, \rm kyr$, as seen from different 
angles defined by the observer’s inclination $\theta_\mathrm{incl}$ (between the line of sight 
and the projection plane of the supernova remnant) and an azimuthal angle $\phi$.  
The panel with $\theta_\mathrm{incl}=0\degree$ and $\phi=0\degree$ shows the overall shape 
of the stellar wind cavity. The termination shock appears brightest at the end of the cavity 
and forms a thick ring around the center of the explosion (Fig.~\ref{fig:emission_maps_0deg}a). 
This feature is described in the 2.5-dimensional context in \citet{meyer_527_mnras_2024}.  
Our 3D treatment of the pulsar wind nebula phase reveals non-axisymmetric features in the 
reverberated termination shock, producing a north–south asymmetry relative to the pulsar’s 
rotational plane, as well as east–west irregularities.  
Fig.~\ref{fig:emission_maps_0deg}b shows the same plerion viewed at 
$\theta_\mathrm{incl}=45\degree$ and $\phi=0\degree$. A rectangular morphology is visible, 
consistent with earlier findings for core-collapse supernova remnants of a 
$35\, \rm M_{\odot}$ star \citep{vanmarle_584_aa_2015,2022MNRAS.515..594M}, and for the 
plerionic case specifically in \citet{meyer_527_mnras_2024}. In our model, a similar rectangular 
shape is observed, with a southern bright extension produced by the polar elongation penetrating 
into the lower part of the cavity.  
The final panel, with $\theta_\mathrm{incl}=90\degree$ and $\phi=0\degree$, shows the 
plerionic core-collapse supernova remnant as a roughly circular structure—an effect of the 
2.5-dimensional nature of the circumstellar medium (Fig.~\ref{fig:emission_maps_0deg}c).

\begin{figure*}
        \centering
        \includegraphics[width=0.47\textwidth]{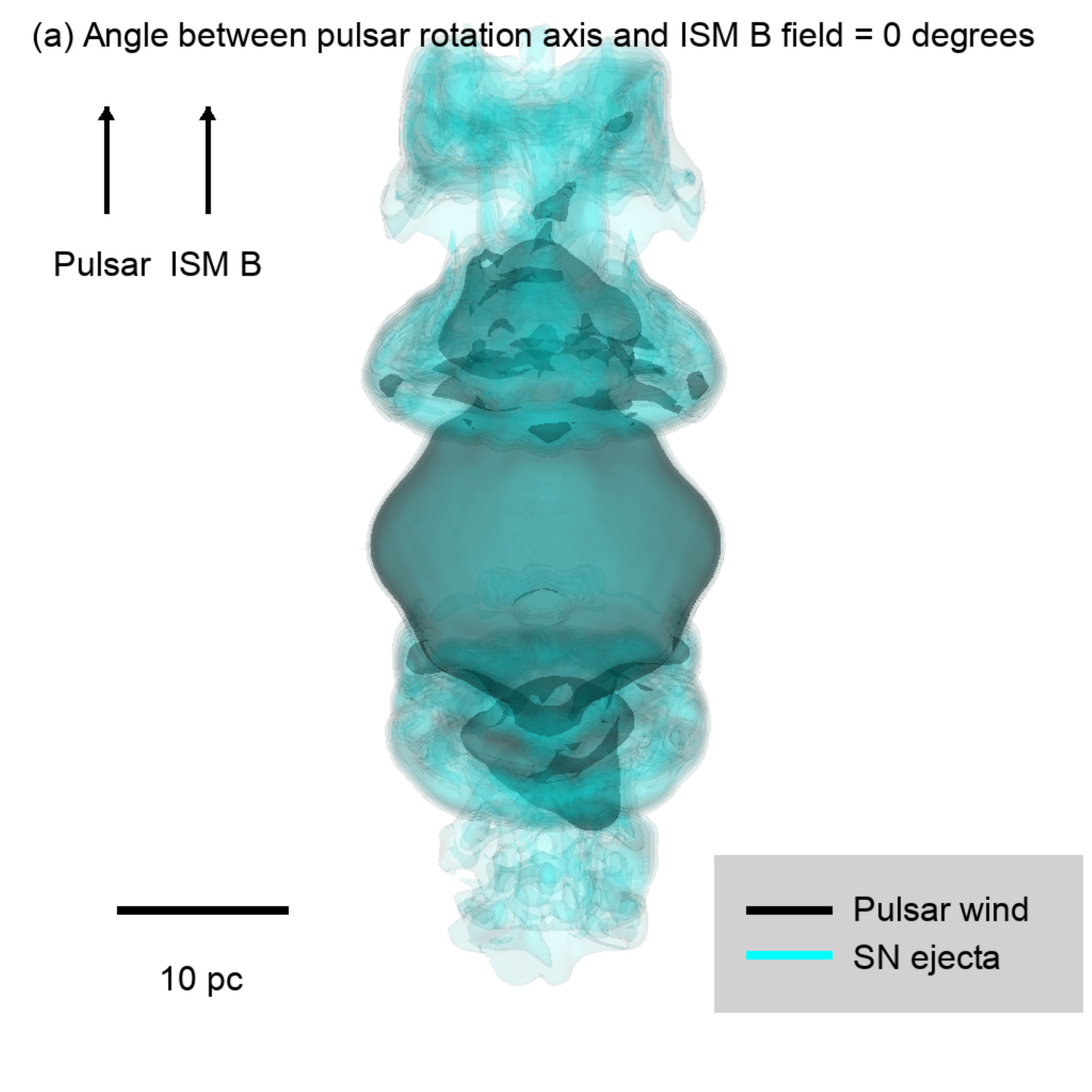}  
        \includegraphics[width=0.47\textwidth]{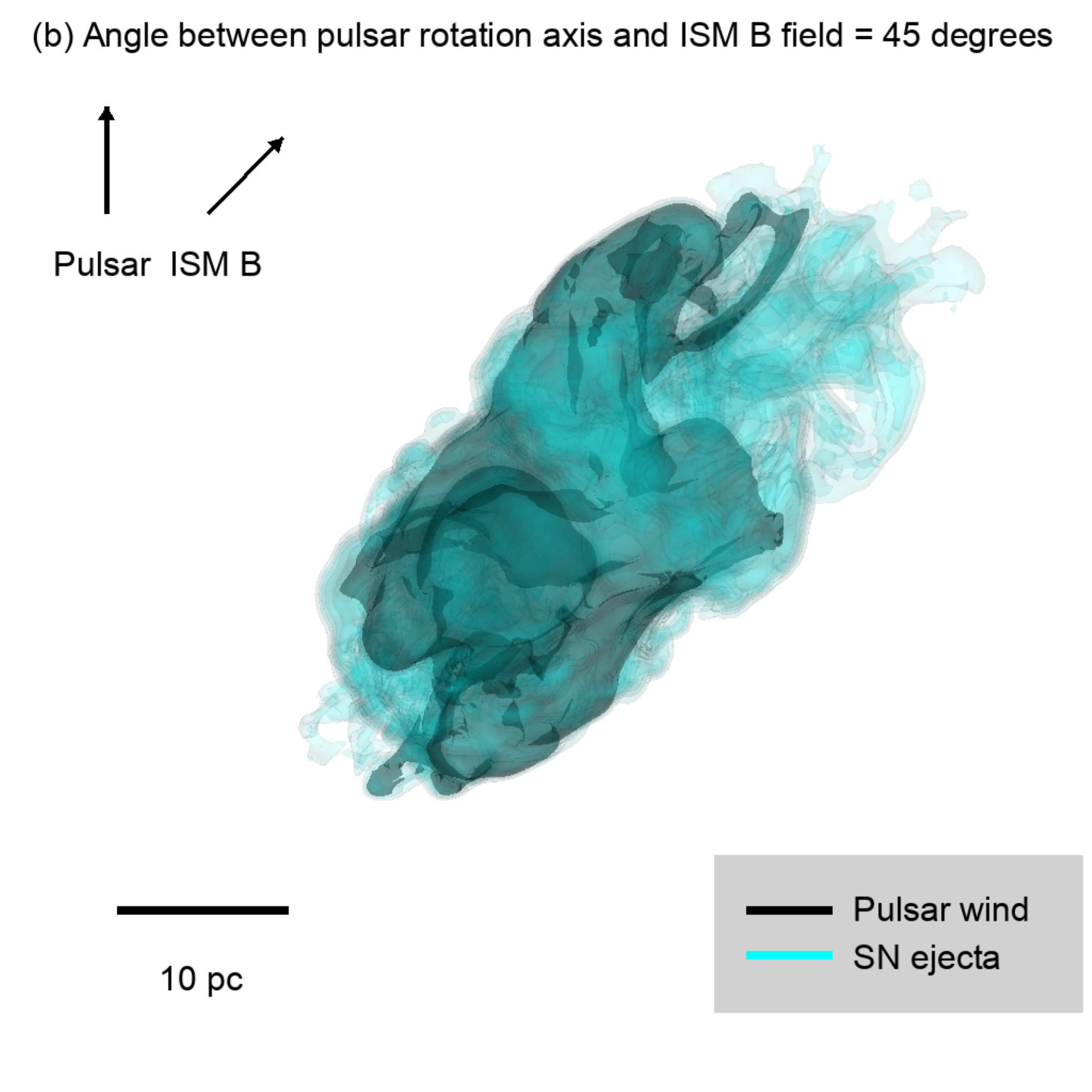}  
        \caption{
        Same as Fig. \ref{fig:3D_PWN_rendering}, displaying the contact discontinuity of the 
        pulsar wind material ($50 \%$ in number density) within the surrounding supernova 
        ejecta (cyan) at time $80\, \rm kyr$ with 
        $\theta_\mathrm{mag}=0\degree$ (a) and $\theta_\mathrm{mag}=45\degree$ (b). 
        \textcolor{black}{
        The figure shows plerion at an advance age of their evolution, when the deviation 
        exerced by the circumstellar medium onto the morphology of the pulsar wind nebula
        is maximum.
        }
        An animated version of this figure is available as online material. 
        }        
        \label{fig:pwn_structure}  
\end{figure*}

Fig.~\ref{fig:emission_maps_30deg}a displays the emission maps of the simulation with 
$\theta_\mathrm{mag}=30\degree$, viewed with $\theta_\mathrm{incl}=0\degree$ and 
$\phi=0\degree$. The peculiar shape of the cavity into which the pulsar wind is channeled 
does not produce a rectangular morphology; instead, it results in an oblong or ovoidal 
structure that is broader in the northern region and narrower in the southern region, 
with an equatorial ring formed by the pulsar wind disc.  
This ring is not visible at $\theta_\mathrm{incl}=45\degree$ and $\phi=0\degree$, 
while the north–south asymmetries in the projected emission become more pronounced, 
further emphasizing the difference in thickness of the pulsar wind nebula’s termination 
shock (Fig.~\ref{fig:emission_maps_30deg}b).  
At $\theta_\mathrm{incl}=90\degree$ and $\phi=0\degree$, the equatorial bulge 
generated by the ring reappears at the center of a cross-shaped morphology of the 
stellar wind cavity (Fig.~\ref{fig:emission_maps_30deg}c).  
In the case of $\theta_\mathrm{mag}=45\degree$, the pulsar wind nebula is seen inside 
the cavity with a distinct equatorial ring and \textcolor{black}{polar elongation}, again exhibiting 
north–south asymmetries due to the fully 3D nature of the simulation 
(Fig.~\ref{fig:emission_maps_45deg}a,~b,~c).  
For $\theta_\mathrm{mag}=90\degree$, the situation is similar, except that the projected 
pulsar wind appears more compact at $\theta_\mathrm{incl}=0\degree$ and 
$\phi=0\degree$ compared to the projections at $\theta_\mathrm{incl}=45\degree$ or 
$\theta_\mathrm{incl}=90\degree$ with $\phi=0\degree$ 
(see Fig.~\ref{fig:emission_maps_90deg}a,~b,~c).

\subsection{Mixing of materials}   
\label{mixing}

\begin{figure}
        \centering
        \includegraphics[width=0.4\textwidth]{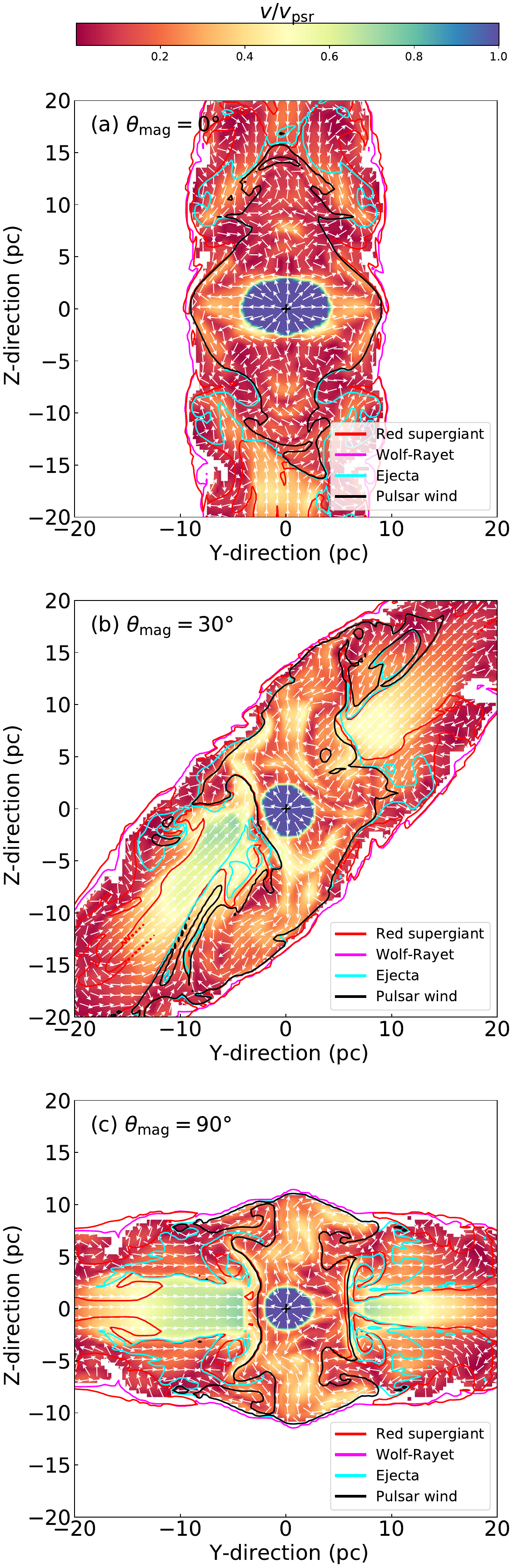}  \\        
        \caption{
Distribution of the velocity field in the 3D supernova remnant models, 
plotted in units of the pulsar wind velocity $v_{\rm psr}$. 
The white arrows trace the velocity field in the yOz plane. 
The figures show the supernova remnants in the $x = 0$ plane of the Cartesian 
coordinate system, at time $70\, \rm kyr$ after the onset of 
the explosion. 
        }
        \label{fig:velocity}  
\end{figure}

\begin{figure}
        \centering
        \includegraphics[width=0.49\textwidth]{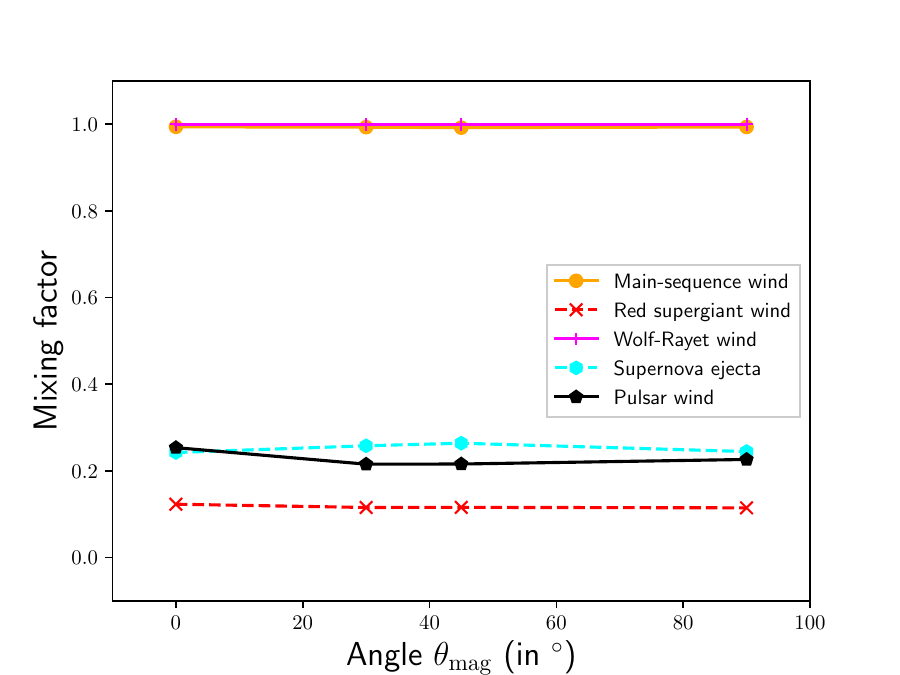}  \\
        \caption{
        Mixing of the different materials in the plerionic supernova remnant at time 
        $80\, \rm kyr$ after the supernova explosion. The line colours distinguish 
        the mixing for the main-sequence (orange), red supergiant (red), Wolf–Rayet 
        (magenta) wind materials, supernova ejecta (cyan), and pulsar wind (black), 
        respectively, displayed as a function of the angle $\theta_\mathrm{mag}$ 
        between the ISM magnetic field and the rotation axis of the pulsar. 
        }
        \label{fig:mixing}  
\end{figure}

Fig.~\ref{fig:mixing} shows the mixing of various materials in the plerionic supernova 
remnant at time $80\, \rm kyr$ after the supernova explosion. The line colors indicate 
the mixing of material from the main-sequence (orange), red supergiant (red), Wolf–Rayet 
(magenta) winds, supernova ejecta (cyan), and pulsar wind (black), respectively, plotted 
as a function of the angle $\theta_\mathrm{mag}$ between the ISM magnetic field and the 
pulsar’s rotation axis.
\textcolor{black}{
The mixing is considered at later times in this study, because we aim at extracting 
conclusions for quasi-steady situations. Before the forward shock of the supernova 
blastwave hits the walls of the cavity and meets the expanding pulsar wind, the 
plerion isessentially growing symmetrically, without any influence from its 
progenitor's circumstellar medium, as described in ~\citep{meyer_mnras_537_2025}. 
}
The mixing efficiency is calculated using passive tracers for each material
(main-sequence, red supergiant, and Wolf–Rayet winds, supernova ejecta, and pulsar wind), 
as well as the mass density and gas velocity. These quantities allow for the tracking of 
advection from the inner boundary, where winds and ejecta are injected, throughout the
supernova remnant, as employed in previous studies~\citep{orlando_aa_444_2005}. Efficiency 
is defined as the ratio of the mass of a given component within the volume of the remnant to the
total mass of the remnant. The spatial extent of the remnant is identified from the
computational domain using a threshold in the density of the unperturbed ISM number.
The figure demonstrates that, at $80\, \rm kyr$ after the explosion, the differences in 
$\theta_\mathrm{mag}$ (the angle between the ISM magnetic field and pulsar spin axis) do 
not fundamentally alter the global mixing within the cavity. By this time, the supernova 
blastwave has had sufficient time to traverse the cavity repeatedly, driven by the 
continuously injected pulsar wind.

The mixing efficiency approaches $100\%$ for both main-sequence and Wolf-Rayet
materials. This can be attributed to two main factors: (i) the large-scale stellar wind
bubble beyond the cavity predominantly contains shocked ISM material due to the absence
of electronic heat conduction in our magnetized circumstellar nebula 
models~\citep{meyer_mnras_464_2017,meyer_mnras_450_2015}, and (ii) all remaining 
main-sequence gas within the cavity has been multiply shocked by successive phases 
of the evolved winds, the supernova ejecta, and the pulsar wind. The Wolf–Rayet wind 
is also highly mixed, as it is the first material encountered by both the blastwave 
and the pulsar wind. 
In contrast, the red supergiant wind, supernova ejecta, and pulsar wind exhibit each a 
mixing efficiency of approximately $20\%$, suggesting a degree of material preservation. 
This indicates the relative purity of the red supergiant wind -rich in hot H-burning products - as well as of the inner remnant material containing Mg, Si, Ca, Ti, and Fe
originating from explosive nucleosynthesis in the supernova and pulsar wind nebulae.%
The quantity and density of materials are different, as well as the time they are 
exposed to other materials. The Wolf-Rayet material is present in smaller quantity 
and is directly shocked by the supernova blastwave. The main-sequence material is 
successively shocked by all materials since the onset of the explosion, while 
supernova ejecta and pulsar wind interact directly. 
These findings are consistent with 2D simulations presented in~\citet{meyer_mnras_537_2025}, 
which examined the mixing of these same species during the first $9\, \rm kyr$ of the 
supernova remnant's evolution inside a stellar wind bow shock.

\subsection{Bending of the polar elongations by the progenitor past stellar wind history}   
\label{bending}

Fig. \ref{fig:pwn_structure} plots the isosurface of the tracer for the pulsar wind 
material at $50\%$ number density in the simulation with $\theta_\mathrm{mag}=0\degree$ 
(Fig. \ref{fig:pwn_structure}a) and with $\theta_\mathrm{mag}=45\degree$ 
(Fig. \ref{fig:pwn_structure}b), respectively. The volume rendering reveals how the 
pulsar wind is channeled into the inner region of the circumstellar medium of the 
progenitor star. 
When the rotation axis of the pulsar is aligned with the local ISM 
magnetic field direction, the pulsar wind nebula grows with the typical morphology of 
an equatorial disc and a \textcolor{black}{polar elongation}, which is itself affected by instabilities. 
In the case with an angle $\theta_\mathrm{mag}=45\degree$, the direction of the stellar 
wind cavity and the direction along which the \textcolor{black}{polar elongation} develops is tilted by an angle approximately equal to $\theta_\mathrm{mag}=45\degree$.

Our results therefore demonstrate a new mechanism to explain how 
the \textcolor{black}{polar elongations} of the pulsar wind nebula grow and 
become unstable. Specifically, the bending of the tails of a 
pulsar wind nebula, which are subject to kink instabilities that self-consistently disturb 
the \textcolor{black}{polar elongation}, has been shown to produce helical \textcolor{black}{polar elongation} deformations~\citep{mignone_mnras_2010}. 
Our findings show that the circumstellar medium of the supernova progenitor plays a 
critical role in shaping the pulsar wind nebula and determining its emission properties, 
as previously demonstrated in the context of static massive stars with a pulsar rotation 
axis aligned with the local ISM magnetic field~\citep{meyer_aa_687_2024}, and in the context 
of runaway pulsars moving through a Cygnus-loop-type supernova remnant created by a runaway 
red supergiant-evolving progenitor~\citep{2025A&A...696L...9M}.
Fig. \ref{fig:velocity} further higlights the velocity field in the yOz plane of the 
supernova remnants a time $70\, \rm kyr$, which is radial inside the free-streaming wind 
region, organised within the disc-like region of the unmixed pulsar wind nebula 
(black contour), and more disorganised in the polar regions of it. Fast bouncing 
material move from the edge of the cavity towards the pulsar wind nebula, imposing 
its bipolar flow to shape as twisted extensions (Fig. \ref{fig:velocity}b).

\section{Conclusion}
\label{conclusion}

This study investigates the influence of the inclination angle between the direction of 
the ISM magnetic field and the rotational axis of the pulsar formed at the explosion 
site of a rotating, magnetized, massive progenitor star with an initial mass of 
$M_\star = 35\, \rm M_{\odot}$. 
The primary objective is to determine how this angular configuration affects the morphology 
of non-thermal radio emission maps produced by synchrotron radiation within the expanding 
supernova remnant. The numerical models employed track the dynamical evolution of the 
supernova blastwave as it propagates through the circumstellar medium of the progenitor 
star, extending up to $35\, \rm kyr$ after the core-collapse event.

To perform these simulations, the circumstellar medium is initially computed using 
2.5D magnetohydrodynamic simulations, which model the stellar evolution from the 
zero-age main-sequence phase to the pre-supernova stage. These models 
incorporate the effects of stellar mass loss through winds and the evolution of the 
large-scale magnetic field in the surrounding environment. The interaction between the 
supernova blastwave and the circumstellar medium is modeled in a multi-step approach, 
beginning with 1D simulations that track the formation and expansion of the supernova 
shock wave as it propagates through the stellar wind. 
These calculations provide insight into the radial structure of the blastwave and the 
development of shocked regions in the circumstellar medium. The subsequent large-scale 
expansion of the supernova remnant is computed within a fully 3D Cartesian numerical 
framework, allowing for the inclusion of asymmetries introduced by the interstellar 
medium's magnetic field and the anisotropic distribution of supernova ejecta. To assess 
the observational implications of these simulations, synchrotron radiation emission 
maps at radio wavelengths are generated.

Following the study of \citet{meyer_aa_687_2024} and \citet{2025A&A...696L...9M}, we found 
that the circumstellar medium of the massive progenitor governs the morphology and the 
emission properties of the pulsar wind nebulae. In this series of simulations, the free parameter 
is the angle between the rotation axis of the pulsar and the local direction 
of the ISM magnetic field, which determines the direction of the cavity formed by the stellar wind 
during the evolutionary history of the massive star. In particular, we found that this angle 
can channel the \textcolor{black}{polar elongation} of the pulsar wind along directions that are not perpendicular 
to the disc of the pulsar wind nebula, offering an alternative explanation to the kink instabilities 
previously invoked to explain the ragged and irregular appearance of these \textcolor{black}{polar elongation} as they propagate 
through the complex region of evolved stellar winds, supernova ejecta, and pulsar wind material.

Emission maps for the synchrotron $1.4\, \rm GHz$ radio waveband reveal a wide variety of 
shapes, combining the projection of the dense, magnetized post-shock region of the stellar 
cavity impacted by the supernova remnant with the projection of the emission generated by 
the pulsar wind material within it. This results in a vast range of morphologies, spanning 
from rectangular to circular forms, as well as more irregular structures exhibiting 
north-south and east-west asymmetries.
The results show that, at the advanced age of the supernova remnants considered, 
the angle between the ISM magnetic field and the pulsar rotation axis does not 
significantly affect the mixing of material within the system.


\begin{acknowledgements} 
The authors acknowledges RES resources provided by BSC in MareNostrum to AECT-2025-1-0004. 
The authors also acknowledge computing time on the high-performance computer "Lise" at 
the NHR Center NHR@ZIB. This center is jointly supported by the Federal Ministry of 
Education and Research and the state governments participating in the NHR 
(www.nhr-verein.de/unsere-partner). 
This work has been supported by the grant PID2024-155316NB-I00 funded by MICIU /AEI /10.13039/501100011033 / FEDER, UE, IEEC APEC2043, and CSIC PIE 202350E189. This work was also supported by the Spanish program Unidad de Excelencia María de Maeztu CEX2020-001058-M and also supported by MCIN with funding from European Union NextGeneration EU (PRTR-C17.I1).
\end{acknowledgements}


\bibliographystyle{aa} 
\bibliography{final_bib_file} 


\clearpage

\onecolumn

\begin{appendix}

\section{Study concept chart}

\begin{figure}[H]
        \centering
        \includegraphics[width=0.95\textwidth]{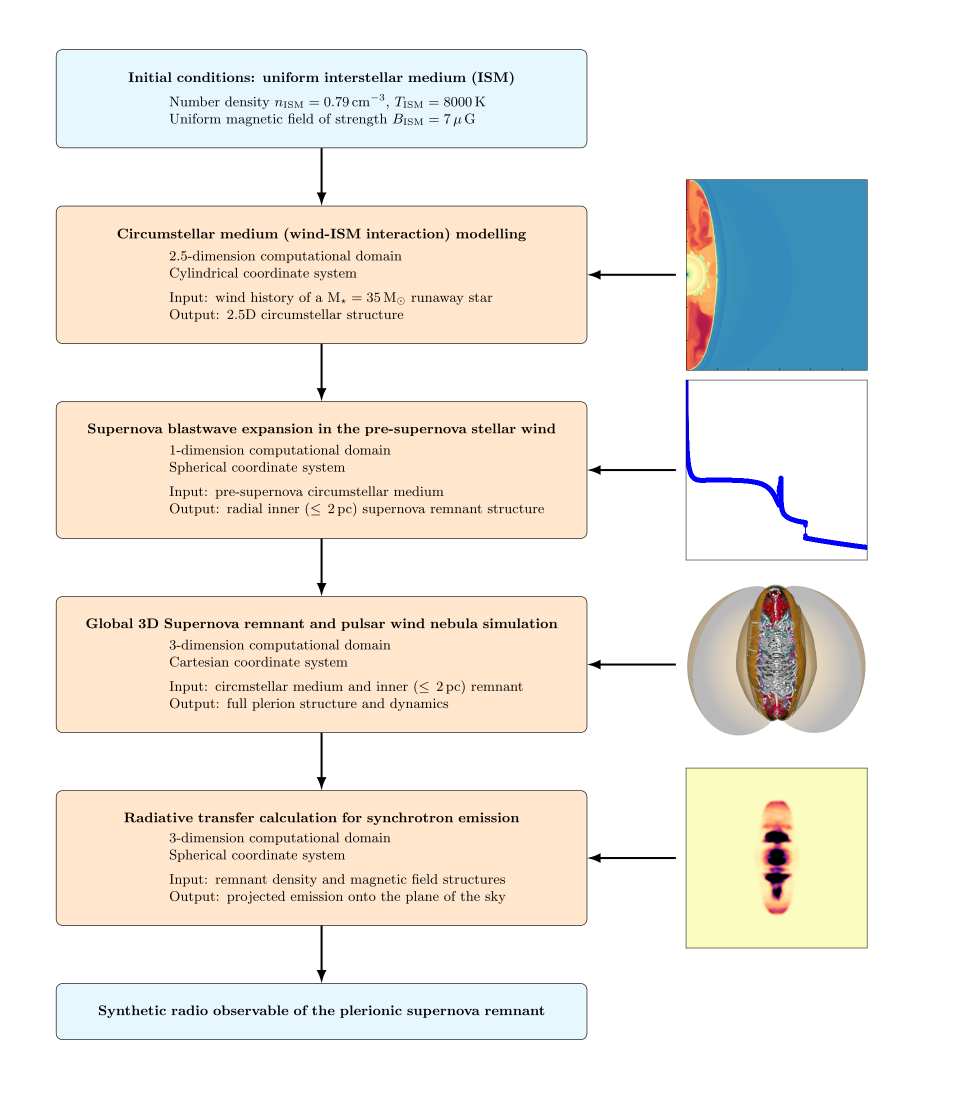}  \\
        \caption{
        \textcolor{black}{
        Concept chart of the workflow adopted in this study. 
        }
        }
        \label{fig:concept_plot}  
\end{figure}


\section{Pulsar wind density field at time $50\, \rm kyr$}

\begin{figure}[H]
        \centering
        \includegraphics[width=0.95\textwidth]{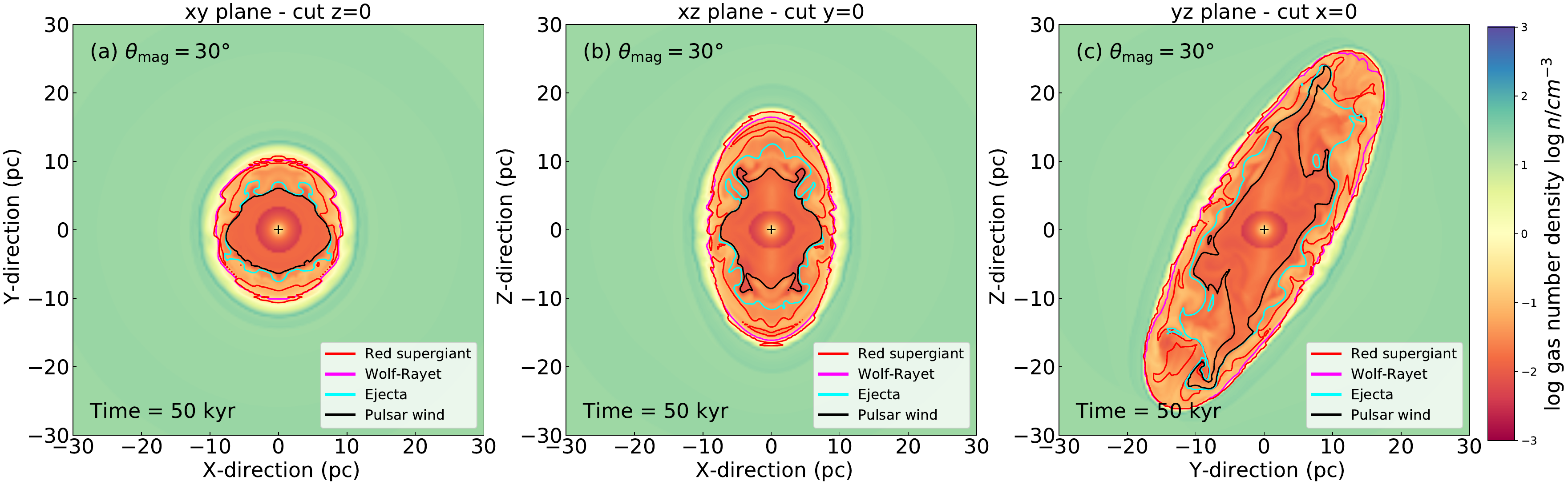}  \\
        \caption{
Same as Fig. \ref{fig:models_20kyr_density_0deg} for $\theta_\mathrm{mag}=30\degree$. 
        }
        \label{fig:models_20kyr_density_30deg}  
\end{figure}

\begin{figure}[H]
        \centering
        \includegraphics[width=0.95\textwidth]{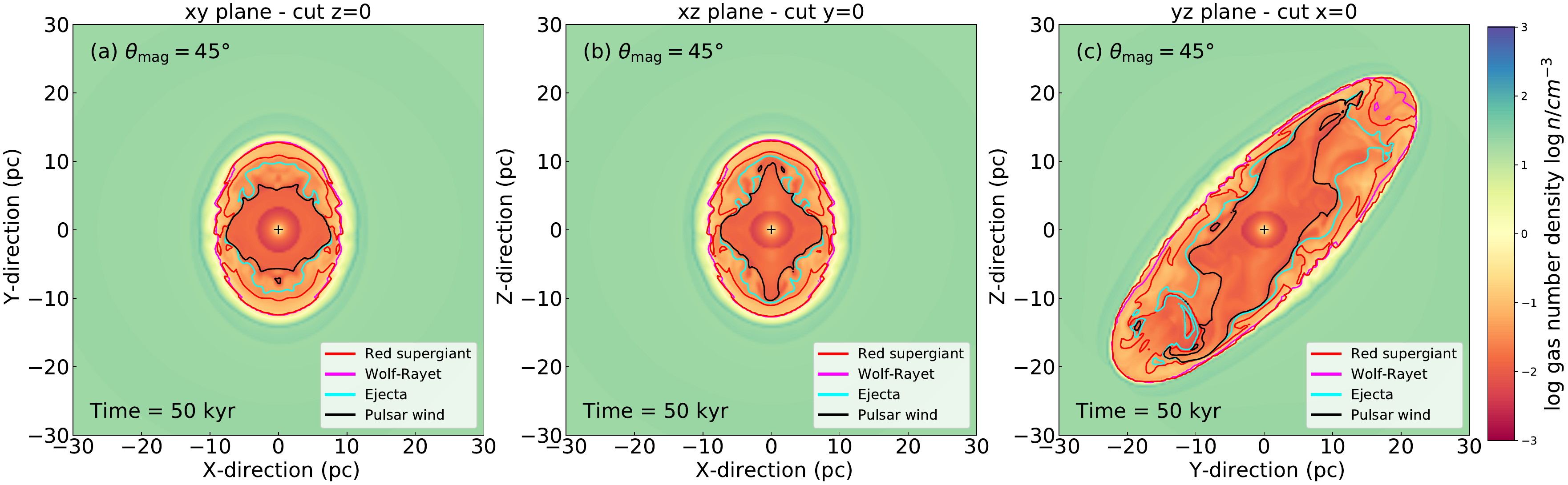}  \\
        \caption{
Same as Fig. \ref{fig:models_20kyr_density_0deg} for $\theta_\mathrm{mag}=45\degree$. 
        }
        \label{fig:models_20kyr_density_45deg}  
\end{figure}

\begin{figure}[H]
        \centering
        \includegraphics[width=0.95\textwidth]{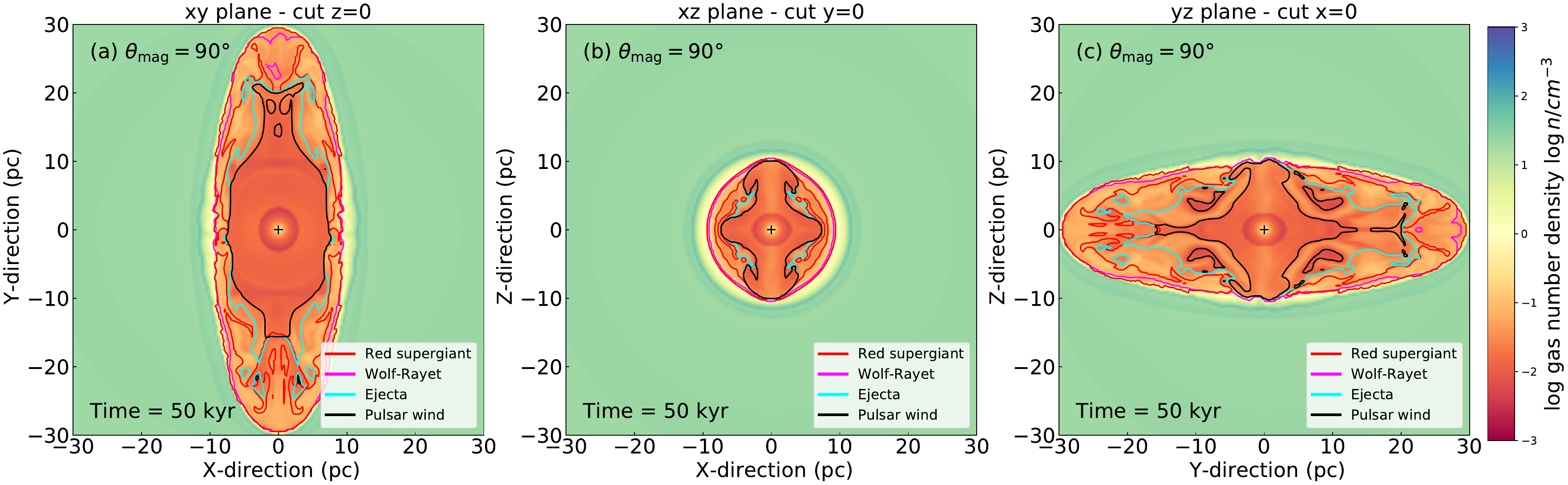}  \\
        \caption{
Same as Fig. \ref{fig:models_20kyr_density_0deg} for $\theta_\mathrm{mag}=90\degree$. 
        }
        \label{fig:models_20kyr_density_90deg}  
\end{figure}

\FloatBarrier

\newpage

\section{Time evolution of the pulsar wind density field}

\begin{figure}[H]
        \centering
        \includegraphics[width=0.95\textwidth]{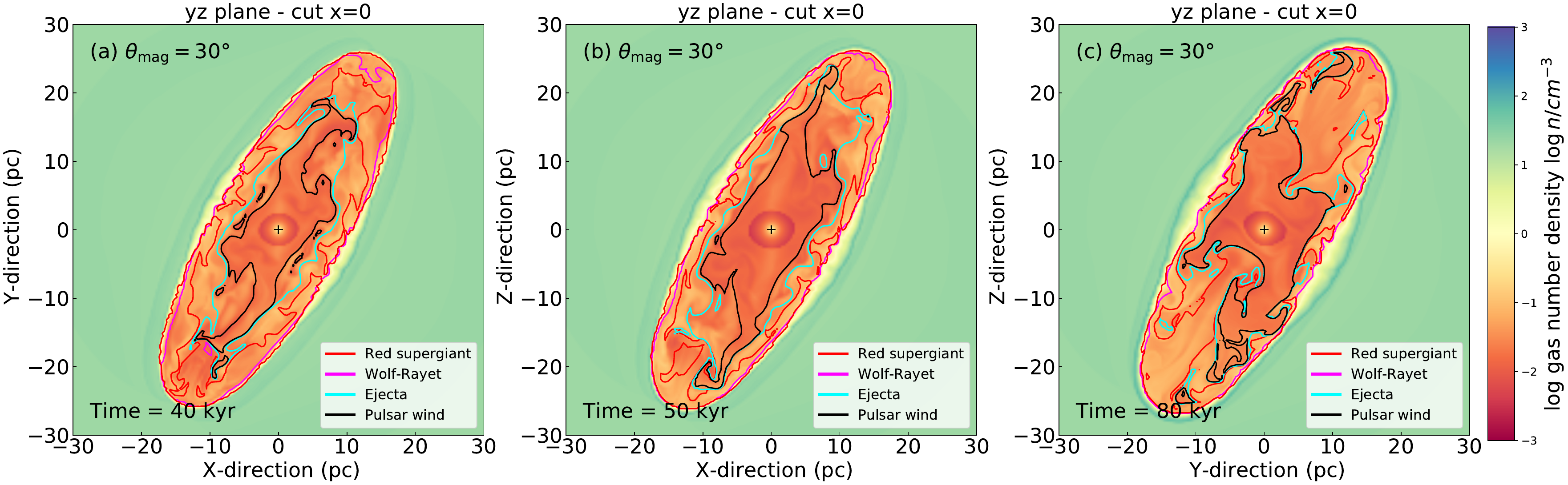}  \\
        \caption{
Same as Fig. \ref{fig:models_30deg_time_evolution} for $\theta_\mathrm{mag}=30\degree$. 
        }
        \label{fig:models_30deg_time_evolution}  
\end{figure}

\begin{figure}[H]
        \centering
        \includegraphics[width=0.95\textwidth]{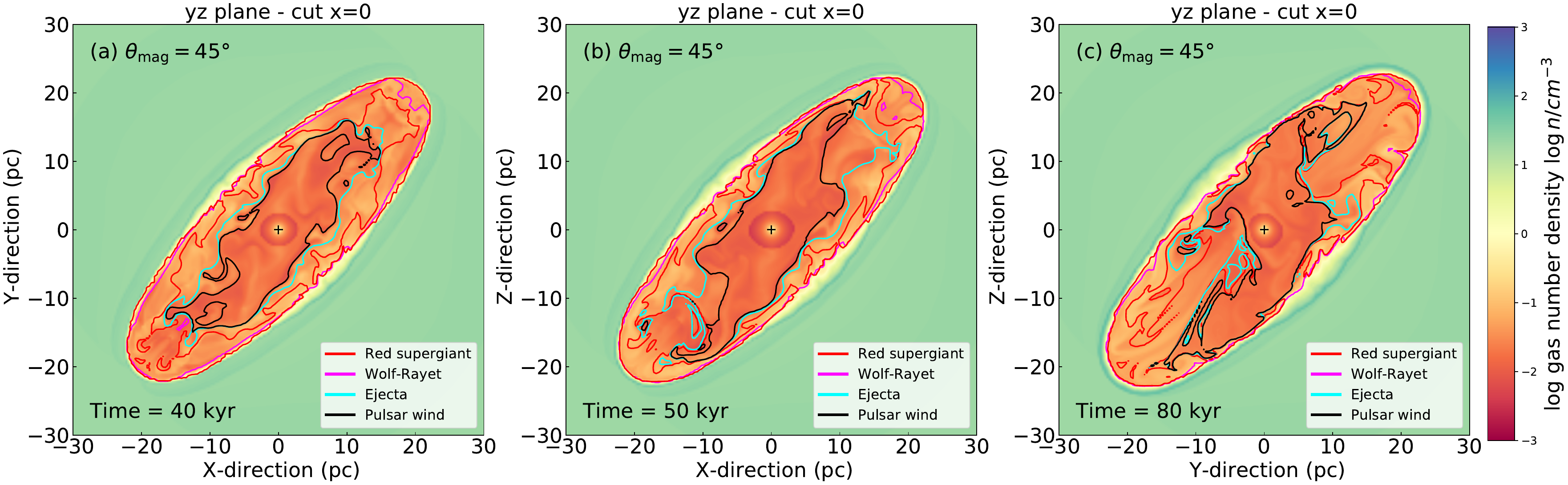}  \\
        \caption{
Same as Fig. \ref{fig:models_45deg_time_evolution} for $\theta_\mathrm{mag}=45\degree$. 
        }
        \label{fig:models_45deg_time_evolution}  
\end{figure}

\begin{figure}[H]
        \centering
        \includegraphics[width=0.95\textwidth]{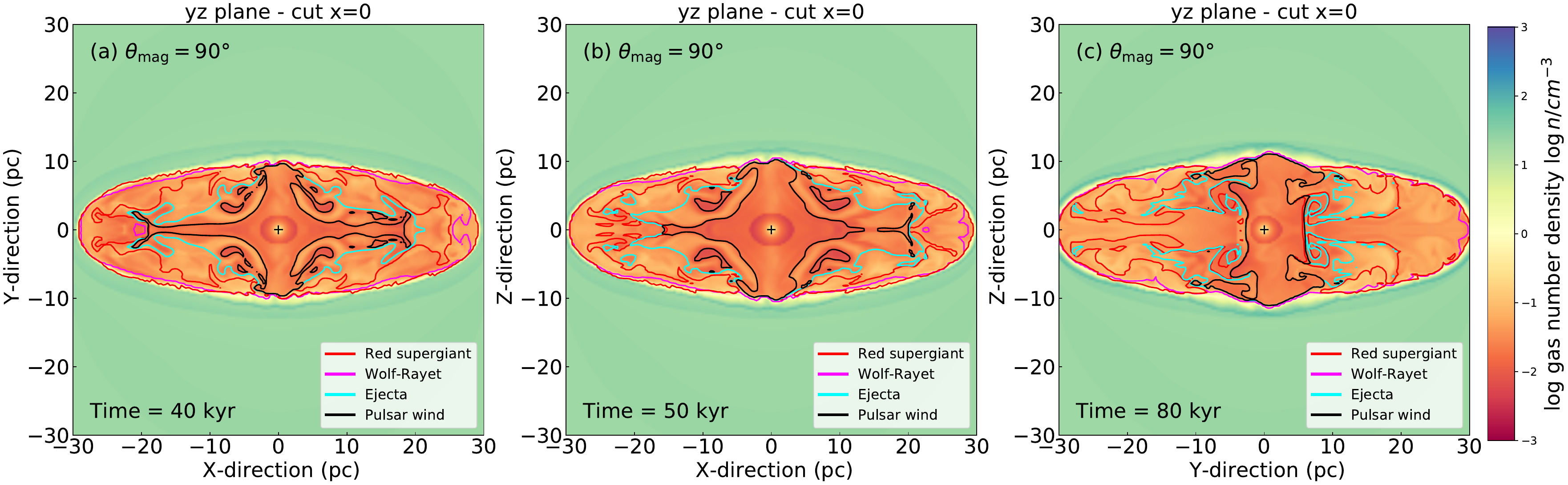}  \\
        \caption{
Same as Fig. \ref{fig:models_90deg_time_evolution} for $\theta_\mathrm{mag}=90\degree$. 
        }
        \label{fig:models_90deg_time_evolution}  
\end{figure}

\FloatBarrier

\section{List of future possible improvements}

\begin{figure}[H]
        \centering
        \includegraphics[width=0.85\textwidth]{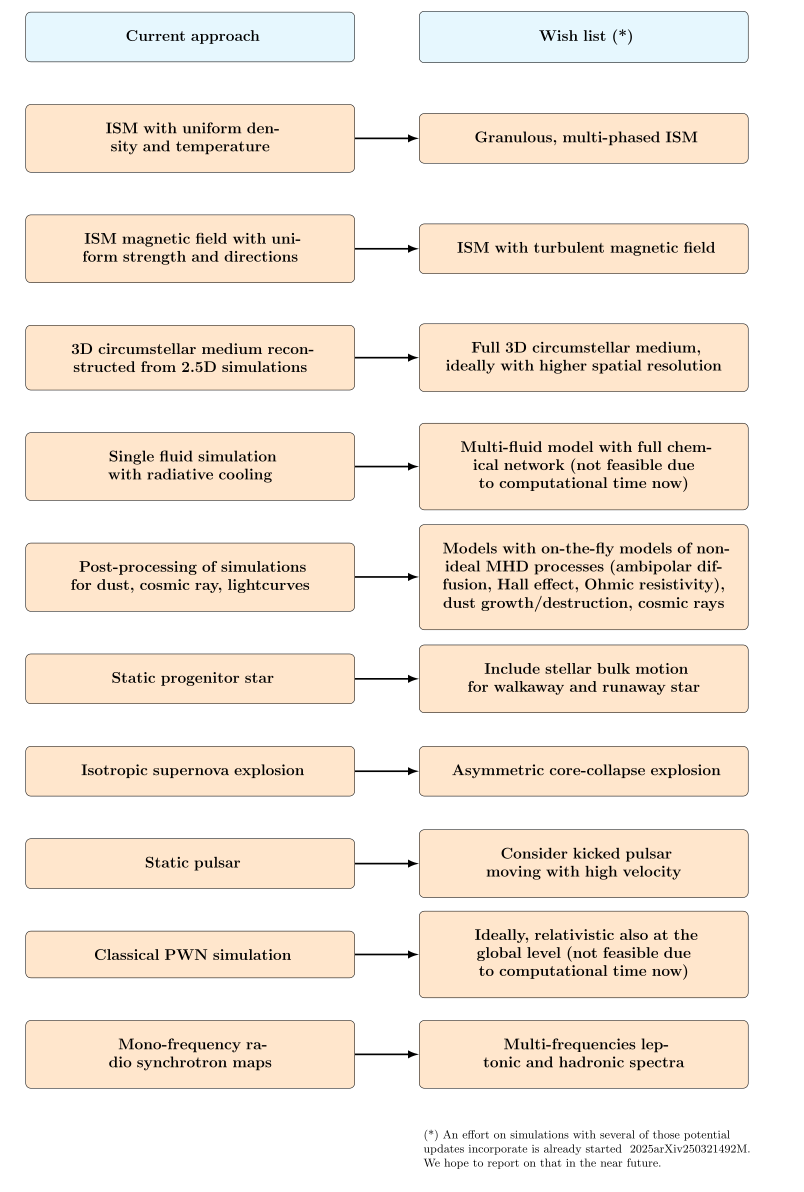}  \\
        \caption{
        Concept chart for future possible improvements of this study. 
        }
        \label{fig:concept_plot_caveats}  
\end{figure}

\FloatBarrier

\section{Emission maps of the pulsar wind nebulae of a static star in a  magnetised ISM}

\begin{figure}[H]
        \centering
        \includegraphics[width=0.85\textwidth]{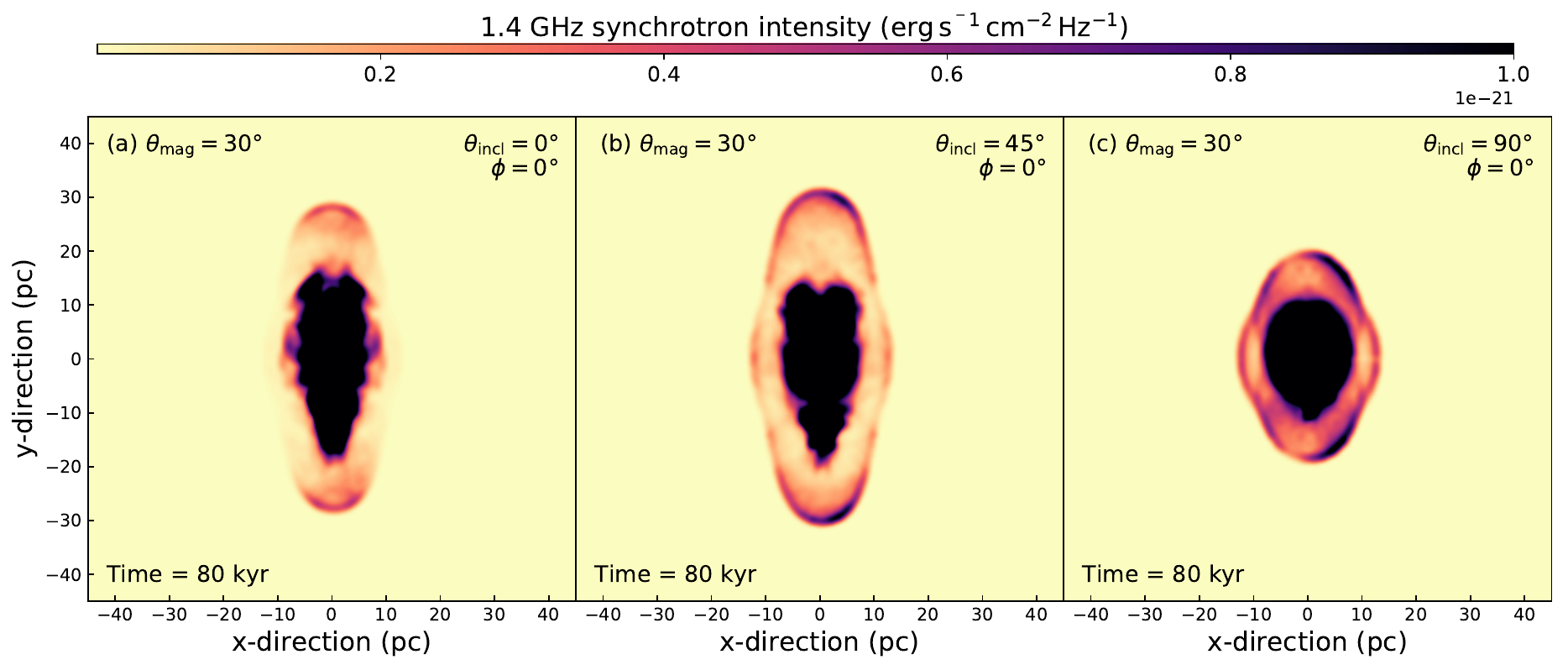}  \\ 
        \caption{
Same as Fig. \ref{fig:emission_maps_30deg} for $\theta_\mathrm{incl}=30\degree$.  
        } 
        \label{fig:emission_maps_30deg}  
\end{figure}v

\begin{figure}[H]
        \centering
        \includegraphics[width=0.85\textwidth]{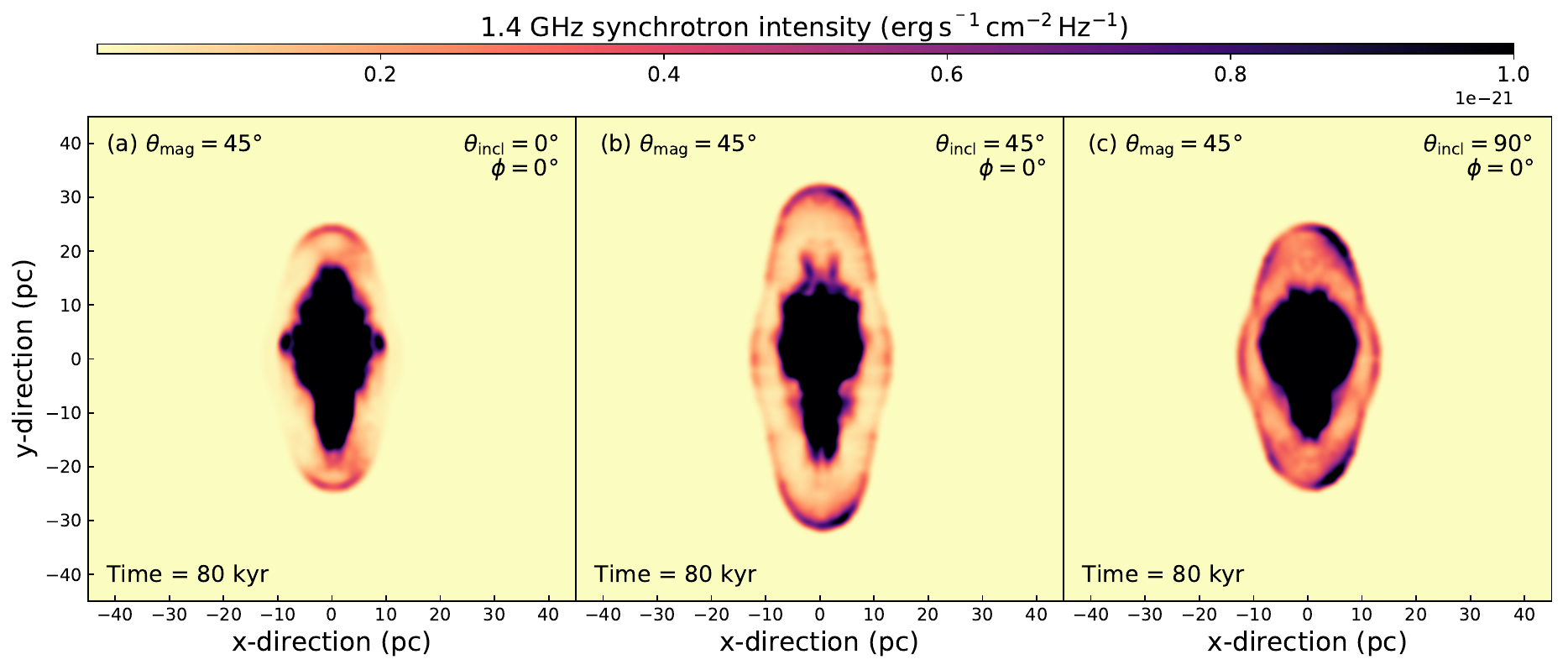}  \\ 
        \caption{
Same as Fig. \ref{fig:emission_maps_30deg} for $\theta_\mathrm{incl}=45\degree$.  
        }
        \label{fig:emission_maps_45deg}  
\end{figure}

\begin{figure}[H]
        \centering
        \includegraphics[width=0.85\textwidth]{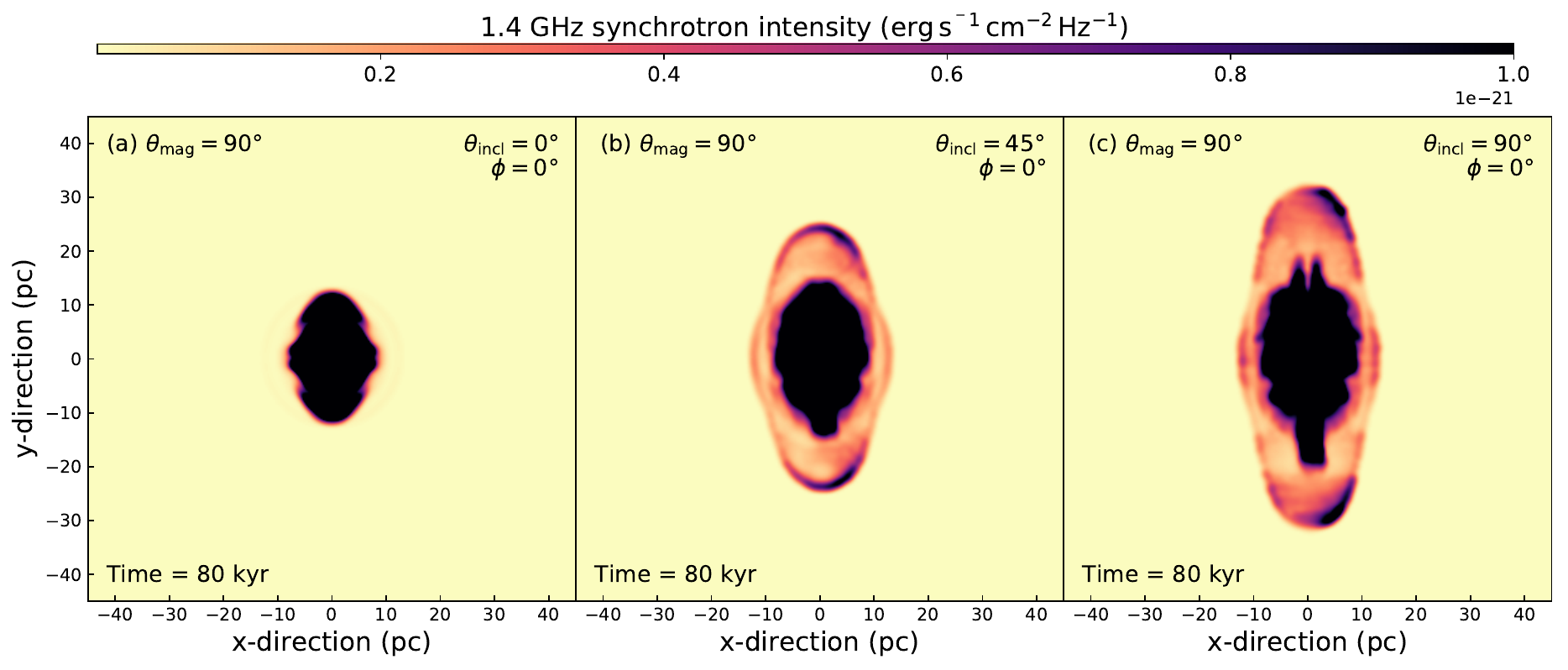} \\  
        \caption{
Same as Fig. \ref{fig:emission_maps_30deg} for $\theta_\mathrm{incl}=90\degree$.   
        }
        \label{fig:emission_maps_90deg}  
\end{figure}

\end{appendix}


\end{document}